\documentclass[english,aps,prb,twocolumn,showpacs]{revtex4-1}
\usepackage[T1]{fontenc}
\usepackage[latin9]{inputenc}
\usepackage{color}
\usepackage{float}
\usepackage{booktabs}
\usepackage{amsmath}
\usepackage{amssymb}
\usepackage{stmaryrd}
\usepackage{graphicx}

\makeatletter

\providecommand{\tabularnewline}{\\}

\definecolor{deColor}{rgb}{0,0.6,0.8}
\definecolor{depink}{rgb}{1,0.05,1}
\usepackage[colorlinks,linkcolor=blue,anchorcolor=blue,citecolor=deColor,urlcolor=blue]{hyperref}

\makeatother

\usepackage{babel}
\begin{document}

\title{Hidden edge Dirac point and robust quantum edge transport in InAs/GaSb
quantum wells}

\author{Chang-An Li}

\affiliation{Department of Physics, The University of Hong Kong, Pokfulam Road,
Hong Kong, China}

\author{Song-Bo Zhang}

\affiliation{Institute of Theoretical Physics and Astrophysics, University of
Würzburg, 97074 Würzburg, Germany}

\author{Shun-Qing Shen}
\email{sshen@hku.hk}

\selectlanguage{english}%

\affiliation{Department of Physics, The University of Hong Kong, Pokfulam Road,
Hong Kong, China\ }

\date{\today }
\begin{abstract}
The robustness of quantum edge transport in InAs/GaSb quantum wells
in the presence of magnetic fields raises an issue on the fate of
topological phases of matter under time-reversal symmetry breaking.
A peculiar band structure evolution in InAs/GaSb quantum wells is
revealed: the electron subbands cross the heavy hole subbands but
anticross the light hole subbands. The topologically protected band
crossing point (Dirac point) of the helical edge states is pulled
to be close to and even buried in the bulk valence bands when the
system is in a deeply inverted regime, which is attributed to the
existence of the light hole subbands. A sizable Zeeman energy gap
verified by the effective g-factors of edge states opens at the Dirac
point by an in-plane or perpendicular magnetic field, however it can
also be hidden in the bulk valance bands. This provides a plausible
explanation for the recent observation on the robustness of quantum
edge transport in InAs/GaSb quantum wells subjected to strong magnetic
fields.
\end{abstract}
\maketitle

\section{Introduction}

The quantum spin Hall (QSH) insulator is a quantum state of matter
with topologically protected helical edge states in the bulk insulating
gap \cite{Kane10rmp,QiXL11rmp,SQS}. The helical edge states will
give rise to the QSH effect which is featured by a quantized conductance
(i.e., $2e^{2}/h$) in the two-terminal measurement at low temperatures
\cite{Kane05prlb}. Theoretically the band crossing point (Dirac point)
of the helical edge states is topologically protected by time-reversal
symmetry, and it opens a minigap once the symmetry is broken (if there
is no other extra symmetry protection). The QSH insulator has been
predicted theoretically \cite{Bernevig06Scien} and confirmed experimentally
in HgTe/CdTe quantum wells \cite{Konig07Scien,Roth09Scien}. Another
promising candidate for QSH insulator is the InAs/GaSb double quantum
well \cite{LiuCX08prl,Knez11prl}. The InAs/GaSb quantum wells possess
a particular electronic phase with inverted band structure, in which
the hybridization of electrons and holes opens a minigap at finite
$k$-vectors, leading to the QSH phase. Due to the mature technology
of material fabrications and potential device applications, there
have been growing efforts exploring the QSH phase in InAs/GaSb quantum
wells \cite{Knez10prb,Knez11prl,Suzuki13prb,Nichele14prl,Mueller15prb,QuF15prl,Nguyen16prl,Karalic16prbR,Nichele17prl}.
Recently, it was observed that the conductance in InAs/GaSb quantum
wells can keep quantized in an in-plane magnetic field up to 12 T
and is insensitive to temperatures ranging from 250 mK to several
Kelvins \cite{DuLJ15prl}. Similar feature was also observed in HgTe/CdTe
quantum wells \cite{Ma15natcomm}. This raises a question about the
fate of the QSH effect under time-reversal symmetry breaking, which
has become a fundamental issue to understand the physics of topological
matter. A number of theoretical efforts have been simulated on this
puzzle \cite{Pikulin14prb,ZhangSB14prb,HuLH16prb}. However the robustness
of the quantized conductance remains poorly understood.

In InAs/GaSb quantum wells, the lowest conduction bands of InAs are
about 150 meV lower than the highest valence bands of GaSb \cite{Altarelli83prb,YangM97prl},
which forms a broken-gap band alignment and leads to the coexistence
of electrons and holes near the charge neutrality point. The application
of gate voltages can shift the band alignment and drive the system
to different electronic phases \cite{Naveh95apl,LiuCX08prl,QuF15prl}.
When the (lowest) electron subbands of InAs lie above the (highest)
heavy hole (HH) subbands of GaSb, the system is in a normal insulator
phase. Whereas the electron subbands lie below the HH subbands, the
system is in an inverted phase and the QSH effect is expected in the
hybridization gap opened by coupling between electron and hole states.
Around the topological phase transition point, the system can be well
described by Bernevig-Hughes-Zhang (BHZ) model which considers four
bands in the lowest energy \cite{Bernevig06Scien,LiuCX08prl}. The
BHZ model, however, fails to explain the robust quantum edge transport
in InAs/GaSb quantum wells in the presence of in-plane magnetic fields,
in which the Dirac point of the helical edge states opens an mini-gap,
leading to the breakdown of quantized conductance. InAs/GaSb quantum
wells could possibly be in a deeply inverted regime where the lower
energy subbands, e.g., the light hole (LH) subbands, will reside above
the electron subbands and may have important influence on the system.
The consideration of the LH subbands may be a resolution to the puzzle.
To this end, re-examination of the band structure of InAs/GaSb quantum
wells and a more comprehensive effective model are needed.

In the present work, a peculiar band structure evolution in InAs/GaSb
quantum wells is revealed when varying the gate voltages. The electron
subbands of InAs can cross the HH subbands of GaSb, and correspondingly
the system transits between a trivial insulator phase and a topological
insulator phase as described by the BHZ model. In contrast, the electron
subbands cannot touch but anticross the LH subbands of GaSb. This
anticrossing behavior does not alter the topology of the system as
no gap closing occurs, however, it may modify the properties of system
near the hybridization gap significantly. We present a six-band effective
model to capture the essential low-energy properties of InAs/GaSb
quantum wells, including the topological phase transition and anticrossing
behavior. One of the key features is that the Dirac point of the edge
states will be pulled to be close to the bulk valence bands when the
electron subbands are lowered to anticross the LH subbands. The application
of a magnetic field, in-plane or perpendicular, opens a sizable Zeeman
energy gap at the Dirac point of the helical edge states, which indicates
the breaking down of the QSH effect. Nevertheless, the energy gap
of \textcolor{black}{edge states could also be hi}dden in the bulk
valence bands up to a large magnetic field, which may account for
recent experimental observation on the robustness of quantum edge
transport under in-plane magnetic fields \cite{DuLJ15prl}. We anticipate
our results can shed some light on experimental observations on the
InAs/GaSb quantum wells and explore novel topological phases of matter
in the future.

The rest of this paper is organized as follows. In Sec. II the band
structure evolution of InAs/GaSb quantum wells is studied, and in
Sec. III a six-band effective model is derived for low-energy physics
of the quantum wells. With the effective model, the properties of
edge states are investigated in Sec. IV. To characterize the response
of the helical edge states to magnetic fields, the effective g-factors
of edge states are calculated in Sec. V. In Sec. VI the robustness
of quantum edge transport under in-plane magnetic fields is addressed
by the numerical calculation of conductance. Finally, Sec. VII contains
the discussions and conclusions.

\section{Band structure evolution of $\mathrm{InAs/GaSb}$ quantum wells }

Both InAs and GaSb have zinc-blende crystal structure and direct gaps
near the $\Gamma$ point, and their low-energy physics can be well
described by the Kane model\ \cite{Kane57JPCS,Winkler}. Considering
the broken-gap band alignment in InAs/GaSb quantum wells and focusing
on the case where the $\Gamma^{6}$ bands of InAs and the $\Gamma^{8}$
bands of GaSb are very close while the $\Gamma^{7}$ bands are far
away in energy and thus can be neglected here. In the basis $\{|\Gamma^{6},1/2\rangle,|\Gamma^{6},-1/2\rangle,|\Gamma^{8},3/2\rangle,|\Gamma^{8},1/2\rangle,|\Gamma^{8},-1/2\rangle,$
$|\Gamma^{8},-3/2\rangle\}$ (Here we use the standard notation that
$|\Gamma^{6},\pm1/2\rangle,$ $|\Gamma^{8},\pm1/2\rangle$ and $|\Gamma^{8},\pm3/2\rangle$
represent the s-like conduction bands, the p-like LH bands, and the
p-like HH bands, respectively), the Kane Hamiltonian for the {[}001{]}
growth direction is given by \cite{Winkler,Novik05prb}
\begin{widetext}
\begin{equation}
H=\left(\begin{array}{cccccc}
T & 0 & -\frac{1}{\sqrt{2}}Pk_{+} & \sqrt{\frac{2}{3}}Pk_{z} & \frac{1}{\sqrt{6}}Pk_{-} & 0\\
0 & T & 0 & -\frac{1}{\sqrt{6}}Pk_{+} & \sqrt{\frac{2}{3}}Pk_{z} & \frac{1}{\sqrt{2}}Pk_{-}\\
-\frac{1}{\sqrt{2}}k_{-}P & 0 & U+V & -\bar{S}_{-} & R & 0\\
\sqrt{\frac{2}{3}}k_{z}P & -\frac{1}{\sqrt{6}}k_{-}P & -\bar{S}_{-}^{\dagger} & U-V & C & R\\
\frac{1}{\sqrt{6}}k_{+}P & \sqrt{\frac{2}{3}}k_{z}P & R^{\dagger} & C^{\dagger} & U-V & \bar{S}_{+}^{\dagger}\\
0 & \frac{1}{\sqrt{2}}k_{+}P & 0 & R^{\dagger} & \bar{S}_{+} & U+V
\end{array}\right),
\end{equation}
\end{widetext}

where
\begin{align}
T & =E_{c}+h'(\gamma_{0}k_{||}^{2}+k_{z}\gamma_{0}k_{z}),\nonumber \\
U & =E_{v}-h'(\gamma_{1}k_{||}^{2}+k_{z}\gamma_{1}k_{z}),\nonumber \\
V & =-h'(\gamma_{2}k_{||}^{2}-2k_{z}\gamma_{2}k_{z}),\nonumber \\
R & =\sqrt{3}h'\gamma_{2}(k_{x}^{2}-k_{y}^{2})-2\sqrt{3}ih'\gamma_{3}k_{x}k_{y},\nonumber \\
\bar{S}_{\pm} & =-\sqrt{3}h'k_{\pm}\left(\{\gamma_{3},k_{z}\}+[\kappa,k_{z}]\right),\nonumber \\
C & =2h'k_{-}[\kappa,k_{z}],
\end{align}
in which ${\bf k}_{\parallel}=(k_{x},k_{y})$, $k_{||}^{2}=k_{x}^{2}+k_{y}^{2}$,
$k_{\pm}=k_{x}\pm ik_{y}$, and $h'=\hbar^{2}/(2m_{0})$. $m_{0}$
is the free electron mass, and $P$ is the Kane momentum matrix element.
$E_{c}$ and $E_{v}$ are the conduction and valence band edges, respectively.
$\gamma_{0,1,2,3}$ and $\kappa$ are the band parameters in the Kane
model. The parameters for InAs, GaSb and AlSb are given in Table \ref{table:parameters}.
We consider the quantum well configuration with InAs and GaSb layers
sandwiched by two AlSb layers at each side along the growth direction
(the $z$-direction). Hence the parameters of the Kane model are spatial
dependent, corresponding to different layers of the quantum wells.
To simulate the experimental setup and for illustration, we will take
12.5 nm InAs/10 nm GaSb with barriers made of 50 nm AlSb at each side
in the quantum well system \cite{DuLJ15prl}.
\begin{widetext}
\begin{table}[H]
\centering

\caption{Parameters in the Kane model for InAs, GaSb and AlSb \cite{Lawaetz71prb,Halvorsen99prb,Nichele17prl}.}

\label{table:parameters}

\begin{tabular}{cccccccccc}
\toprule
  & \qquad{}$E_{g}${[}eV{]}\quad{} & \quad{}$P${[}eV$\cdot\mathrm{\textrm{Å}}${]}\qquad{} & \enskip{}$\gamma_{1}$\qquad{} & \qquad{}$\gamma_{2}$\qquad{} & \qquad{}$\gamma_{3}$\qquad{} & \qquad{}$\gamma_{0}$\qquad{} & \qquad{}$\kappa$\qquad{} & \quad{}$E_{c}${[}eV{]}\quad{} & \quad{}$E_{v}${[}eV){]}\quad{}\tabularnewline
\midrule
InAs & 0.41 & 9.19 & 19.67 & 8.37 & 9.29 & 1/0.03 & 7.68 & -0.15 & -0.56\tabularnewline
\midrule
GaSb & 0.8128 & 9.23 & 11.8 & 4.03 & 5.26 & 1/0.042 & 3.18 & 0.8128 & 0\tabularnewline
\midrule
AlSb & 2.32 & 8.43 & 4.15 & 1.01 & 1.75 & 1/0.18 & 0.31 & 1.94 & -0.38\tabularnewline
\bottomrule
\end{tabular}
\end{table}

\begin{table}[H]
\centering

\caption{Parameters in the six-band effective model for $V_{0}$=-100meV, $L_{\text{InAs}}=12.5$
nm, $L_{\text{GaSb}}=10$ nm and $L_{\text{AlSb}}=50$ nm.}
\label{parameter-effectivemodel}

\begin{tabular}{ccccccccccc}
\toprule
Parameters  & \enskip{}$B_{e}${[}eV$\cdot\mathrm{\mathring{A}}^{2}${]}\  & \ $B_{h}${[}eV$\cdot\mathrm{\mathring{A}}^{2}${]}\  & \ $B_{l}${[}eV$\cdot\mathrm{\mathring{A}}^{2}${]}\  & \ $D_{el}${[}eV$\cdot\mathrm{\mathring{A}}^{2}${]}\  & \ $P_{eh}${[}eV$\mathrm{\cdot\textrm{Å}}${]}\  & \ $P_{el}${[}eV$\cdot\mathrm{\textrm{Å}}${]}\  & \ $P_{lh}${[}eV$\mathrm{\cdot\textrm{Å}}${]}\  & \ $P_{e}${[}eV$\cdot\mathrm{\textrm{Å}}${]}\  & \ $P_{l}${[}eV$\mathrm{\cdot\textrm{Å}}${]}\  & \tabularnewline
\midrule
Value  & 81.3 & -31.2 & -60 & 40 & 0.45 & 0.11 & \textcolor{black}{0.61} & 0.13 & 0.29 & \tabularnewline
\midrule
Parameters & $\gamma_{2eh}$ & $\gamma_{3eh}$ & $\gamma_{2lh}$ & $\gamma_{3lh}$ & $Q_{e}${[}eV{]} & $Q_{l}${[}eV{]} & $Q_{el}${[}eV{]} & $E_{e}${[}eV{]} & $E_{h}${[}eV{]} & $E_{l}${[}eV{]}\tabularnewline
\midrule
Value  & 1.88 & 2.45 & -3.3 & -4.3 & 0.76 & 0.21 & 0.38 & -0.0283 & -0.0115 & \textendash 0.0529\tabularnewline
\bottomrule
\end{tabular}
\end{table}
\end{widetext}

We assume the confinement effect in the $z$-direction and replace
the operator $k_{z}$ with $-i\partial_{z}$ in the Hamiltonian. The
full Hamiltonian of the quantum wells takes the form
\begin{equation}
H_{\text{full}}=H_{K}(k_{x},k_{y},-i\partial_{z})+V(z).\label{eq:Hamiltonian1}
\end{equation}
Here $V(z)$ is the confinement potential and is also spatial dependent.
The subbands dispersions and corresponding eigenstates are obtained
by solving the Schrödinger equation:
\begin{equation}
H_{\text{full}}|\Psi^{\xi}(k_{x},k_{y},z)\rangle=E^{\xi}|\Psi^{\xi}(k_{x},k_{y},z)\rangle,
\end{equation}
where $\xi$ is the subband index, and $|\Psi^{\xi}(k_{x},k_{y},z)\rangle=\exp[ik_{x}x+ik_{y}y]F^{\xi}(z)$
with $F^{\xi}(z)$ an envelope function. The envelope function approximation
can be employed to solve the eigen problem of the quantum wells \cite{LiJ09prb}.
$F^{\xi}(z)$ can be expanded in terms of plane waves
\begin{equation}
F^{\xi}(z)=\sum_{\lambda=1}^{6}\sum_{n=-N}^{N}\frac{1}{\sqrt{L}}a_{n,\lambda}^{\xi}e^{ik_{n}z}|\lambda\rangle,
\end{equation}
where $k_{n}=2\pi n/L$ with $n=0,\pm1,\pm2,\cdots,\pm N$ ($N$ is
a positive integer), and $L=L_{\mathrm{InAs}}+L_{\mathrm{GaSb}}+2L_{\mathrm{AlSb}}$
is the total width of InAs/GaSb quantum wells. $a_{n,\lambda}^{\xi}$
are the corresponding expansion coefficients. Here we use $|\lambda\rangle$
($\lambda$ = 1, 2, $\cdots$ , 6) to denote the basis set of wave
functions where $|1\rangle$ and $|2\rangle$ are for $|\Gamma^{6},\pm1/2\rangle$,
$|3\rangle$ and $|6\rangle$ are for $|\Gamma^{8},\pm3/2\rangle$,
and $|4\rangle$ and $|5\rangle$ are for $|\Gamma^{8},\pm1/2\rangle$.
For the numerical calculations, we take $N=30$ which is accurate
enough for the low-energy physics.

\begin{figure}[h]
\includegraphics[width=8.5cm]{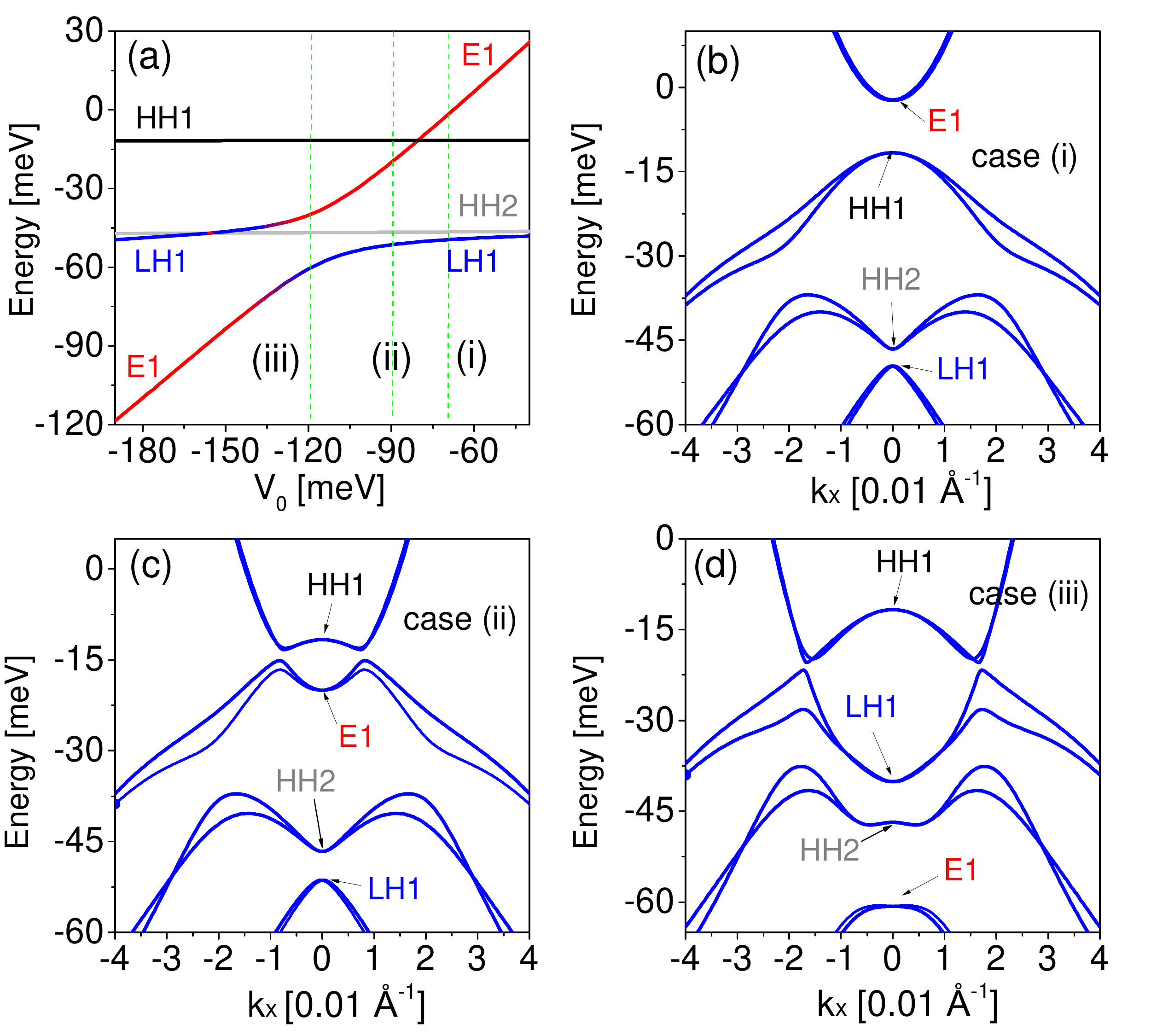}

\caption{The energy spectrum of InAs/GaSb quantum wells in different phases.
(a) The energies of the lowest-energy subbands at the $\Gamma$ point
as functions of the broken gap $V_{0}$. The band structures at $k_{y}=0$
with (b) $V_{0}=-70$ meV; (c) $V_{0}=-90$ meV; (d) $V_{0}=-120$
meV.}
\label{fig:Band-structure}
\end{figure}

Different electronic phases can be realized by varying the broken
gap $V_{0}$, the energy difference of band edges between the $\Gamma^{6}$
bands of InAs and the $\Gamma^{8}$ bands of GaSb, which is supposed
to be tunable by gate voltages \cite{LiuCX08prl,QuF15prl}. Figure
\ref{fig:Band-structure}(a) shows the energies of the lowest energy
subbands at the $\Gamma$ point as functions of $V_{0}$. One can
see that when decreasing $V_{0}$, the lowest electron ($E1$) subbands
cross the highest HH ($HH1$) subbands, showing a topological phase
transition. For a large $V_{0}(>-80$ meV), the system is a trivial
insulator as shown in Fig. \ref{fig:Band-structure}(b) and should
not possess robust edge states, which is labeled as case (i). For
a smaller $V_{0}(<-80$ meV), the system transfers from the trivial
insulating phase to a shallowly inverted phase labeled by case (ii).
A hybridization gap will open at the crossing point, as shown in Fig.\ \ref{fig:Band-structure}(c),
and the QSH effect is expected \cite{LiuCX08prl}. The low-energy
properties of the system near the phase transition point $V_{0}(\sim-80$
meV) can be well described by the BHZ model \cite{Bernevig06Scien}.
Decreasing $V_{0}$ further, the $E1$ subbands does not touch but
anticross the highest LH ($LH1$) subbands. We label the deeply inverted
phase after the anticrossing as case (iii). The transition from cases
(ii) to (iii) is topologically trivial since there is no gap closing,
however, some important properties (e.g., the property of edge states)
near the system gap are changed, as will be shown below. The corresponding
band structure for case (iii) is presented in Fig. \ref{fig:Band-structure}(d),
which exhibits giant spin-orbit splitting close to the hybridization
gap. The spin-orbit splitting due to the structure inversion asymmetry
may lead to fully spin polarized states \cite{Nichele17prl}.

\section{Six-band effective model}

The topologically non-trivial band structure indicates the existence
of helical edge states across bulk insulating gap with the open boundaries
according to the bulk-edge correspondence \cite{Hatsugai93prl,Qi06prb,Graf2013}.
To find the helical edge states and investigate the low-energy properties
of InAs/GaSb quantum wells, it is helpful to derive an effective model,
just as the BHZ model \cite{Bernevig06Scien}. Noting that without
gate voltage the InAs/GaSb quantum wells tend to stay in the deeply
inverted phase of case (iii), the $LH1$ subbands may have significant
influence on the system and thus should also be considered. A six-band
effective model which involves the $E1$, $HH1$ and $LH1$ subbands
can be constructed, following a similar procedure of Refs.\ \cite{Bernevig06Scien,Roth10njp}.

Generally the full bulk Hamiltonian can be split into two parts
\begin{equation}
H_{\text{full}}=H_{0}({\bf k}_{\parallel}=0,-i\partial_{z},z)+H'({\bf k}_{\parallel},-i\partial_{z},z),\label{eq:Full Hamiltonian}
\end{equation}
where $H_{0}$ describes the system at the $\Gamma$ point (i.e.,
${\bf k}_{\parallel}=0)$  and $H'$ can be treated as a perturbation
around the $\Gamma$ point. First, we can numerically solve the Schrödinger
equation $H_{0}|\Psi_{0}^{\xi}\rangle=E_{0}^{\xi}|\Psi_{0}^{\xi}\rangle$
and obtain the eigenenergies $E_{0}^{\xi}$ and the corresponding
eigenstates $|\Psi_{0}^{\xi}\rangle$. The Hamiltonian $H_{0}$ is
effectively decoupled to four blocks: the electron subbands couple
only with the LH subbands, while the HH subbands decouple from them.
We can treat these decoupled blocks separately. Three eigen wave functions
with components of the $E1$ and $HH1$ bands, or of the $LH1$ subbands
can be written as
\begin{alignat}{1}
\langle z|E1,+\rangle & =\left(\psi_{e1}(z),0,0,\psi_{e4}(z),0,0\right)^{T},\\
\langle z|HH1,+\rangle & =\left(0,0,\psi_{h3}(z),0,0,0\right)^{T},\\
\langle z|LH1,+\rangle & =\left(\psi_{l1}(z),0,0,\psi_{l4}(z),0,0\right)^{T},
\end{alignat}
where $T$ means transpose. The envelope function components $\psi_{e(h,l)}(z)$
can be found by expanding the eigenstates in terms of plane waves,
as introduced previously. Carrying out the time-reversal operation
on the above wave functions, we have other three eigen wave functions
\begin{alignat}{1}
\langle z|E1,-\rangle & =\left(0,\psi_{e1}^{*}(z),0,0,-\psi_{e4}^{*}(z),0\right)^{T},\\
\langle z|HH1,-\rangle & =\left(0,0,0,0,0,\psi_{h3}^{*}(z)\right)^{T},\\
\langle z|LH1,-\rangle & =\left(0,-\psi_{l1}^{*}(z),0,0,\psi_{l4}^{*}(z),0\right)^{T}.
\end{alignat}

Next, with the six lowest energy states at the $\Gamma$ point as
a basis set, we can project the Hamiltonian (\ref{eq:Full Hamiltonian})
and obtain a two-dimensional six-band effective model. In the ordered
basis $\left\{ |E1,+\rangle,|E1,-\rangle,\right.$$|HH1,+\rangle,$
$|LH1,+\rangle,\left.|LH1,-\rangle,|HH1,-\rangle\right\} $, the effective
Hamiltonian reads
\begin{align}
H & ({\bf k}_{\parallel})=H_{0}({\bf k}_{\parallel})+\delta H,\label{eq:effect Hamiltonian}
\end{align}
\begin{align}
H_{0}({\bf k}_{\parallel}) & =\left(\begin{array}{cccccc}
T_{e} & -\frac{P_{e}k_{-}}{\sqrt{6}} & -\frac{P_{eh}k_{+}}{\sqrt{2}} & Dk^{2} & \frac{P_{el}k_{-}}{\sqrt{6}} & R_{eh}\\
-\frac{P_{e}k_{+}}{\sqrt{6}} & T_{e} & -R_{eh}^{\dagger} & -\frac{P_{el}k_{+}}{\sqrt{6}} & -Dk^{2} & \frac{P_{eh}k_{-}}{\sqrt{2}}\\
-\frac{P_{eh}k_{-}}{\sqrt{2}} & -R_{eh} & T_{h} & -\frac{P_{lh}k_{-}}{\sqrt{2}} & R_{lh} & 0\\
Dk^{2} & -\frac{P_{el}k_{-}}{\sqrt{6}} & -\frac{P_{lh}k_{+}}{\sqrt{2}} & T_{l} & \frac{P_{l}k_{-}}{\sqrt{6}} & R_{lh}\\
\frac{P_{el}k_{+}}{\sqrt{6}} & -Dk^{2} & R_{lh}^{\dagger} & \frac{P_{l}k_{+}}{\sqrt{6}} & T_{l} & -\frac{P_{lh}k_{-}}{\sqrt{2}}\\
R_{eh}^{\dagger} & \frac{P_{eh}k_{+}}{\sqrt{2}} & 0 & R_{lh}^{\dagger} & -\frac{P_{lh}k_{+}}{\sqrt{2}} & T_{h}
\end{array}\right),\\
\delta H & =\Delta V\left(\begin{array}{cccccc}
Q_{e} & 0 & 0 & Q_{el} & 0 & 0\\
0 & Q_{e} & 0 & 0 & -Q_{el} & 0\\
0 & 0 & 0 & 0 & 0 & 0\\
Q_{el} & 0 & 0 & Q_{l} & 0 & 0\\
0 & -Q_{el} & 0 & 0 & Q_{l} & 0\\
0 & 0 & 0 & 0 & 0 & 0
\end{array}\right)
\end{align}
where $k_{\pm}=k_{x}\pm ik_{y}$, $T_{e(l,h)}=E_{e(l,h)}+B_{e(l,h)}k_{\parallel}^{2}$,
and $R_{e(l)h}=\sqrt{3}h'\gamma_{2e(l)h}(k_{x}^{2}-k_{y}^{2})-i2\sqrt{3}h'\gamma{}_{3e(l)h}k_{x}k_{y}$.
Here and after, we choose a fixed broken gap $V_{0}$ as reference
and take $\Delta V$ as a variation from $V_{0}$ to tune the band
structure evolution for convenience. $H_{0}({\bf k}_{\parallel})$
describes the system with the broken gap $V_{0}$. $\delta H$ is
the modification by $\Delta V$, the small change of the broken gap,
it shows clearly how the whole band structure varies as tuning gate
voltages. The diagonal terms $Q_{e}$ and $Q_{l}$ in $\Delta H$
will shift the position of $E1$ and $LH1$ subbands, as shown in
Fig.\ \ref{fig:Band-structure}(a). There is no diagonal term for
the $HH1$ subbands in $\Delta H$, which is consistent with Fig.\ \ref{fig:Band-structure}(a)
in which the $HH1$ subbands nearly do not shift. The off-diagonal
term $Q_{el}$ is crucial for the anticrossing behavior. It couples
the $E1$ and $LH1$ subbands even at the $\Gamma(k_{x}=k_{y}=0)$
point, preventing them from touching with each other. In this way,
the effective model not only covers the physics of the BHZ model but
also captures the anticrossing behavior of the energy bands. The parameters
in this effective Hamiltonian can be found straightforwardly in the
projection, and they depend on the details of the quantum wells (i.e.,
the thickness of the quantum wells and the broken gap reference $V_{0}$,
etc.). For the considered quantum well configuration (i.e., the thickness
of 50/12.5/10/50 nm for AlSb/InAs/GaSb/AlSb), the parameters in the
effect model are provided in Table \ref{parameter-effectivemodel}.

\begin{figure}[h]
\includegraphics[width=8cm]{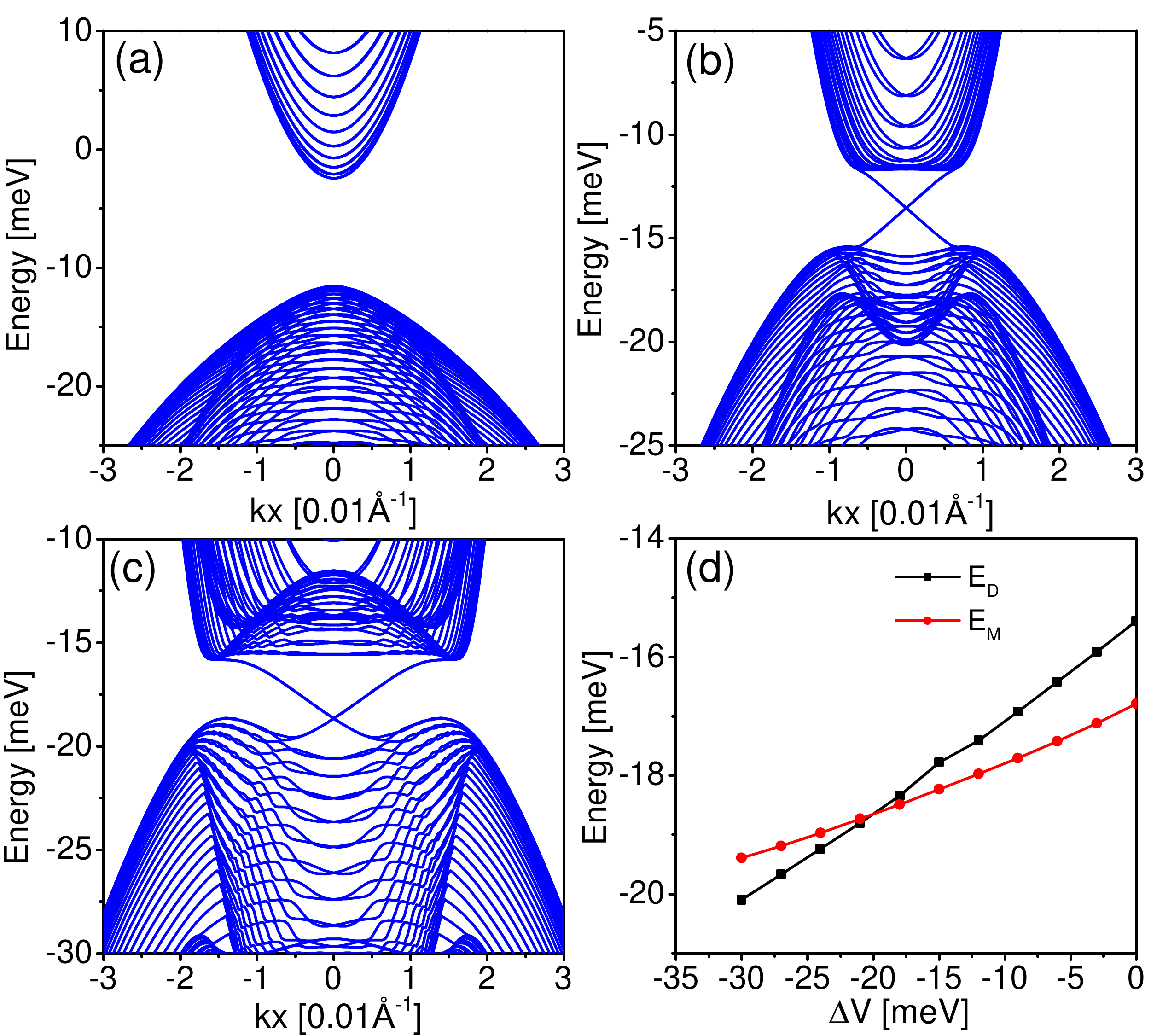}

\caption{Energy spectrum of bulk and edge states of the system with periodic
and open boundaries in the $x$ and $y$ direction, respectively.
(a) for $\Delta V=30$ meV , (b) for $\Delta V=10$ meV, (c) for $\Delta V=-20$
meV, and (d) the energy position of Dirac point ($E_{D}$) and maximum
point of valence bands ($E_{M}$) as function of $\Delta V$. $V_{0}=-100$
meV is taken for all figures. }
\label{fig:TB bands}
\end{figure}

\section{Hidden Dirac point of the helical edge states}

With the six-band effective model, we are in a position to investigate
the energy dispersions of the edge states for the topologically non-trivial
cases (ii) and (iii). This can be accomplished numerically by means
of tight-binding calculations. The tight-binding model can be obtained
by discretizing the effective Hamiltonian Eq. (\ref{eq:effect Hamiltonian})
on a square lattice. In the long wavelength limit, we use the approximation
$k_{i}\approx\sin(k_{i}a)/a$ and $k_{i}^{2}\approx2[1-\cos(k_{i}a)]/a^{2}$
with $i=x,\ y$ and $a$ the lattice constant. We take $a=20\mathring{\mathrm{A}}$
which is a good approximation to the continuum limit. To find the
edge states solution, we apply the open boundary condition along the
$y$ direction while the periodic boundary condition along the $x$
direction. Thus $k_{x}$ remains a good quantum number and the system
is diagonal in $k_{x}$.

Figures\ \ref{fig:TB bands}(a)-(c) plot the energy spectrum of the
effective model in the absence of external fields, corresponding to
the cases (i)-(iii) as mentioned above. For the trivial insulator
case (i), there is a direct system gap and no edge dispersion as shown
in Fig.\ \ref{fig:TB bands}(a). In both cases (ii) and (iii) as
shown in Figs.\ \ref{fig:TB bands}(b) and (c), there are two pairs
of gapless and doubly degenerate helical edge bands across the bulk
insulating gap, as expected for the QSH effect. Nevertheless, for
case (iii) the Dirac point of the helical edge states is close to
and even ``buried'' by the bulk valence states, which is in contrast
to case (ii) where the Dirac point is well exposed in the middle of
the bulk gap {[}see Fig. \ref{fig:TB bands}(b){]}. As reducing $\Delta V$
further, the Dirac point $E_{D}$ approaches the maximum point of
the bulk valence bands $E_{M}$, and eventually it is hidden by the
bulk valence bands, as shown in Fig. \ref{fig:TB bands}(d).

The hidden Dirac point of edge states in case (iii) can be attributed
to the anticrossing between $E1$ and $LH1$ subbands by comparing
with Fig. \ref{fig:Band-structure}(a). We find that the Dirac point
can be hidden only around the value of $\Delta V$ where the anticrossing
behavior occurs. The Dirac point will not be buried in the bulk states
but well exposed in the bulk gap if the $LH1$ subbands are not taken
into account. The hidden Dirac point is also related to the strong
anisotropy in the system, which inherits from the bulk Kane model.
The finding that the Dirac point of edge states can be hidden in the
bulk bands serves as the basis for the robust quantum edge transport
in InAs/GaSb quantum wells under time-reversal breaking as will be
discussed in the following, and it is one of our main results.

\section{Effective $g$-factors of edge states}

A magnetic field ${\bf B}$ breaks time-reversal symmetry, and consequently
the Dirac point of the edge states will no longer be topologically
protected if there is no other hidden symmetry. The time-reversal
symmetry breaking can be evidenced by a gap opening in the helical
edge states, which originates from the Zeeman and the orbital coupling
effects of the bulk electrons in an external magnetic field. In the
six-band effective model, the Zeeman term can be written as
\begin{equation}
H_{Z}=H_{c}^{Z}\varoplus H_{v}^{Z},
\end{equation}
with
\begin{equation}
H_{c}^{Z}=(1/2)g_{e}\mu_{B}{\bf s}\cdot{\bf B},
\end{equation}
for electrons in the s-like $E1$ bands, and
\begin{equation}
H_{v}^{Z}=g_{h}\mu_{B}{\bf J}\cdot{\bf B}.
\end{equation}
for the p-like $HH1$ and $LH1$ bands \cite{Winkler,Beugeling12prb}.
Here ${\bf s}=\left\{ s_{x},s_{y},s_{z}\right\} $ are the Pauli matrices
for spin 1/2, ${\bf J}$ are the $4\times4$ angular momentum matrices
for $j=3/2$, and $\mu_{B}$ is the Bohr magneton. $g_{e}$ and $g_{h}$
are the g-factors for bulk electrons and holes, respectively, and
are taken to be $g_{e}$$=-10.0$ and $g_{h}$$=0.3$ \cite{ChaoKA03jn,MuXY16apl}
in the following.

\begin{figure}[h]
\includegraphics[width=8cm]{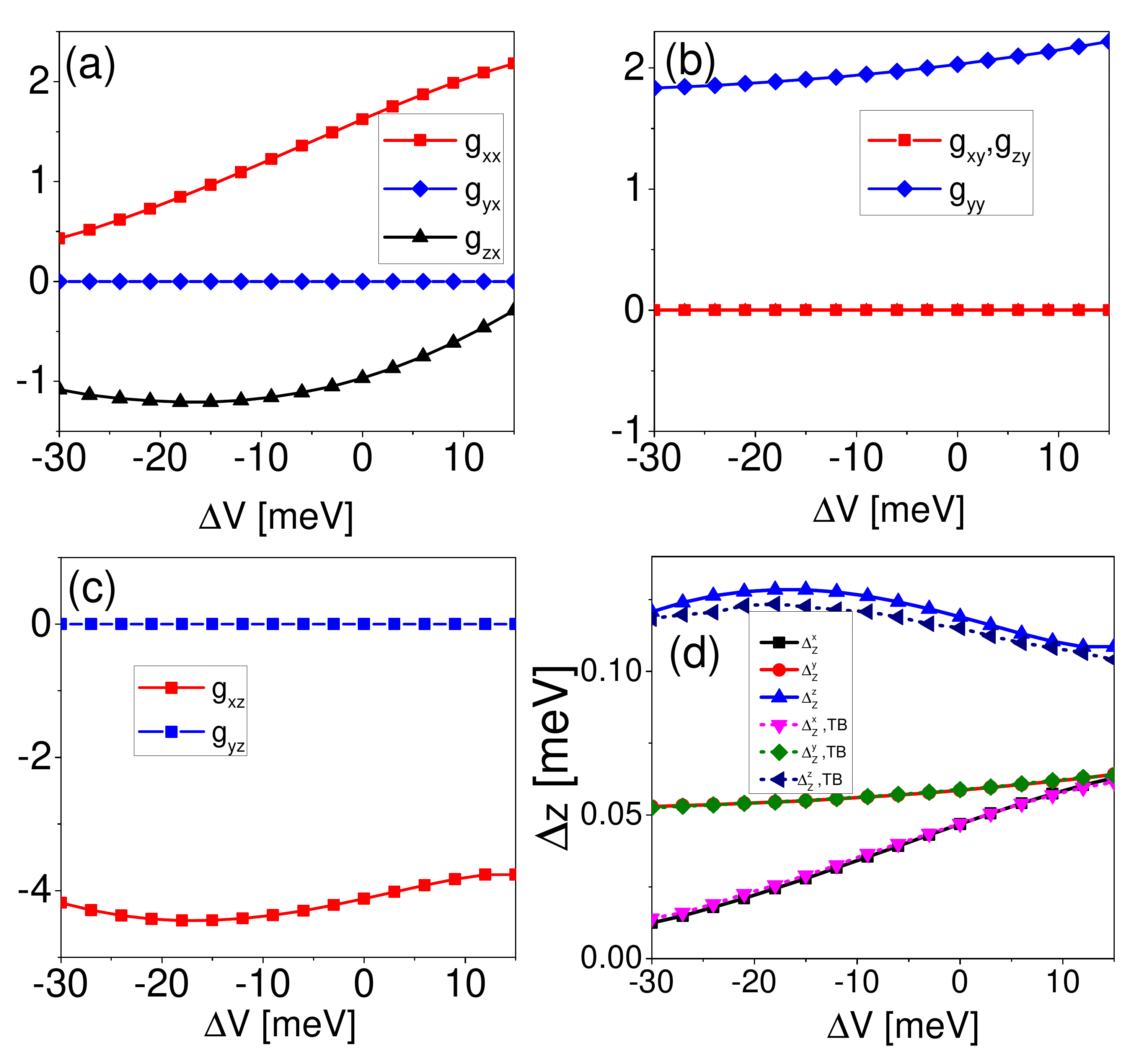}

\caption{Effective g-factors of edge states and Zeeman energy gaps for edge
states spectrum. The effective g-factor tensor elements of the edge
states as function of $\Delta V$ for magnetic field along (a) $x$,
(b) $y$, and (c) $z$ direction, respectively. (d) The energy gaps
$\Delta_{Z}^{x,y,z}$ of edge states opened by three principal magnetic
field of 0.5 T as functions of $\Delta V$. $V_{0}=-100$ meV is taken
for all figures. \label{fig:g-factors}}
\end{figure}

The response of the helical edge states to the magnetic fields can
be examined by projecting the Zeeman term in the space spanned by
the two helical edge states $|\psi_{0+}\rangle$ and $|\psi_{0-}\rangle$
at the $\Gamma$ point. Note that $|\psi_{0+}\rangle$ and $|\psi_{0-}\rangle$
are time-reversal to each other. The corresponding effective Zeeman
coupling can be summarized as
\begin{equation}
\mathcal{H}_{\text{edge}}^{Z}=\dfrac{\mu_{B}}{2}\sum_{i,j=x,y,z}g_{ij}\sigma_{i}B_{j},\label{eq:Zeeman}
\end{equation}
where the $g_{ij}$ is the effective g-factor tensor and $\sigma_{x,y,z}$
are the Pauli matrices for the edge states space. Reminding that the
effective model for the helical edge states takes the form $\mathcal{H}_{\text{edge}}^{0}=\hbar v_{F}k_{x}\sigma_{z}$
where $v_{F}$ is the effective velocity.

The g-factor tensor is attributed to the the fact that the two helical
edge states at the $\Gamma$ point are not the eigenstates of electron
spin. Figures \ref{fig:g-factors}(a,b,c) plot the values of the g-factor
elements $g_{ij}$ for different $\Delta V$, from which several points
are worthy addressing. For a perpendicular magnetic field $B_{z}$,
considering the contribution from the orbital angular momentum coupling
to  $B_{z}$, a large value of $g_{zz}$ is obtained. This large $g_{zz}$
just shifts the position of the degeneracy (Dirac) point of the helical
edge states in the $k_{x}$ direction, whereas it does not open an
energy gap (so we do not show it here). However, a non-zero $g_{xz}$
does open an energy gap. Here the Peierls substitution is performed
as $t^{ij}\rightarrow t^{ij}\exp[\frac{2\pi i}{\phi_{0}}\int_{i}^{j}d{\bf \boldsymbol{\ell}}\cdot{\bf A}]$
where $\phi_{0}=h/e$ is the magnetic flux quantum, and $t^{ij}$
is the hopping integral between sites $i$ and $j$. For an in-plane
field, the orbital contribution to g-factors is ignorable as electrons
are confined in the quantum wells. $g_{xx}$ and $g_{yy}$ always
take non-zero values, which indicates that an in-plane magnetic field
also opens a gap in the edge states. These values of Zeeman gap calculated
from the effective g-factor tensor of edge states match well with
those obtained directly from the spectrum {[}see Fig.\ \ref{fig:g-factors}(d){]}.
Therefore the non-zero g-factors indicate an opened gap at the Dirac
point of helical edge states under time-reversal symmetry breaking
\cite{YangY11prl}, and the QSH effect is broken down. It is also
interesting to find that the effective g-factors of edge states show
an evident anisotropy. Especially for the in-plane magnetic fields,
though both edge Zeeman gaps $\Delta_{\text{Z }}^{x,y}$ decay as
decreasing $\Delta V$, $\Delta_{\text{Z }}^{x}$ decays much faster,
which indicates that the anisotropy is enhanced for a small $\Delta V$.
Finally, we note that $\Delta_{\text{Z }}^{x,y}$ can reach the order
of 1 meV for a magnetic field of 10 T, which are experimentally measurable
at low temperatures. However, these Zeeman gap $\Delta_{Z}^{x,y}$
could be hidden since the Dirac point would be hidden by the bulk
valence bands after the anticrossing behavior at a small $\Delta V$.

\section{Robustness of the quantum edge transport}

Now let us address the robustness of the edge transport in the InAs/GaSb
quantum wells in the inverted regime. It is known that the quantized
two-terminal conductance $2e^{2}/h$ of a QSH insulator  is a consequence
of the helical edge states, which has been measured experimentally
in the InAs/GaSb quantum wells. Unexpectedly under in-plane magnetic
fields either along or normal to the boundary the quantized conductance
value remains quantized for mesoscopic samples and persists up to
12 T \cite{DuLJ15prl}. To understand the robustness of quantized
conductance plateau, the evolution of the band structure subjected
to an in-plane external magnetic field has been explored. The in-plane
magnetic field effect can be included by considering that the InAs
and GaSb layers are spatially separated\ \cite{YangM97prl,QuF15prl,HuLH16prb}.
An in-plane magnetic field applied along the open boundary $B_{y}$
will not only open an energy gap at the Dirac point of the edge states,
but also tilt the bulk energy spectra and reduce the bulk gap \cite{QuF15prl,HuLH16prb}.
Henceforth, there is no direct gap between the edge states and the
valence bands if the Dirac point is buried in the bulk. Similar effect
happens for the in-plane magnetic field $B_{x}$ normal to the open
boundary.

\begin{figure}[h]
\includegraphics[width=8.5cm]{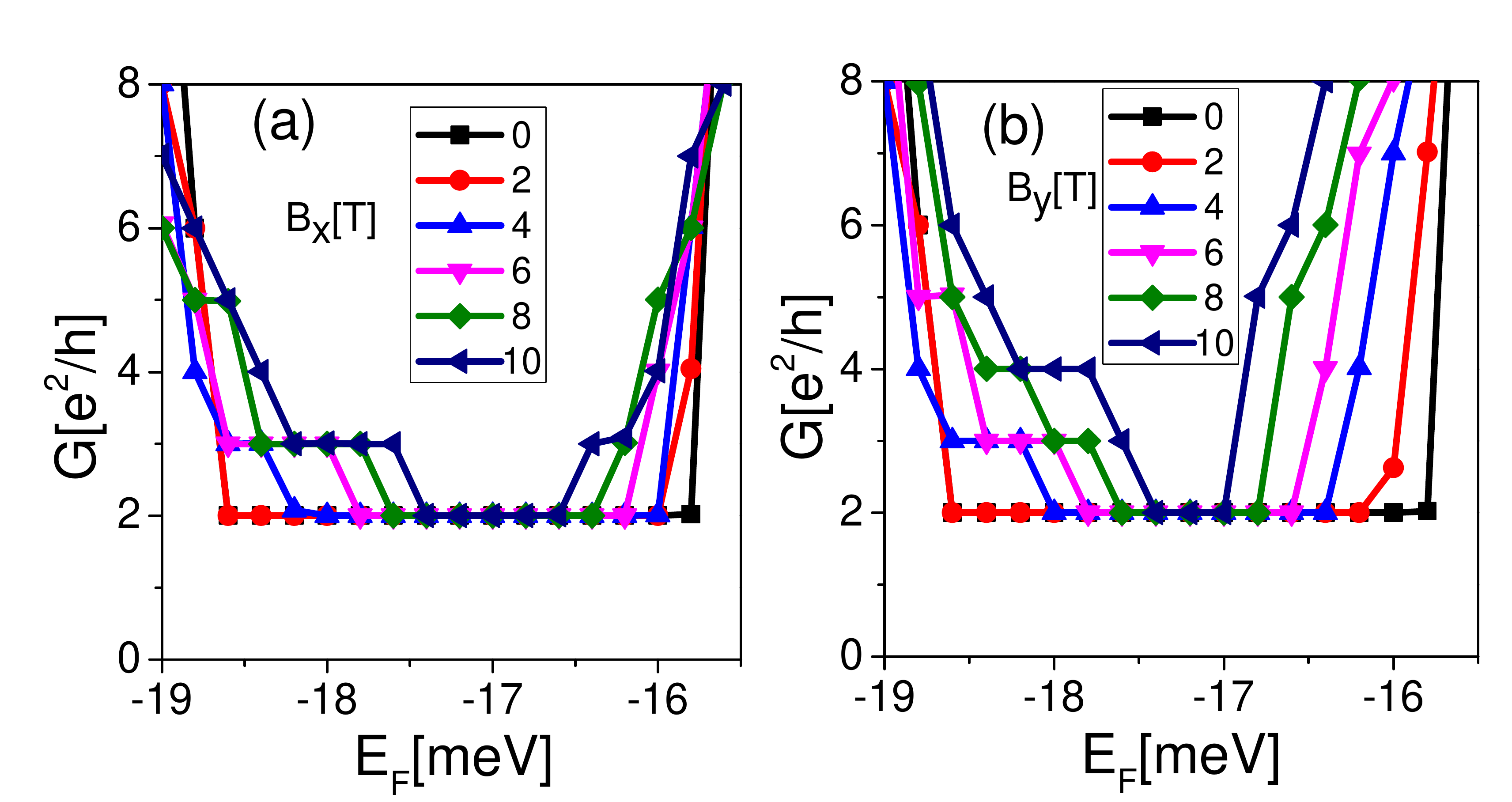}

\caption{The two-terminal conductance $G$ of InAs/GaSb quantum wells in magnetic
fields. (a) $G$ as a function of the Fermi energy $E_{F}$ in the
presence of different magnetic field $B_{x}$ along the $x$ direction.
(b) the same as (a) but with the magnetic fields applied along the
$y$ direction. $\Delta V=-20$ meV and $V_{0}=-100$ meV are taken
for both figures. }
\label{fig:conductance}
\end{figure}

Consider a ribbon geometry of the InAs/GaSb quantum wells. The two-terminal
conductance is calculated as a function of the Fermi energy $E_{F}$
under different in-plane magnetic fields by means of the Landauer-Büttiker
formalism in a clean sample. The sample geometry considered consists
of a rectangular central region (size $L_{x}\times L_{y}=200a\times150a$)
and two semi-infinite leads are connected to it as source and drain
leads. With the help of recursive Green's function technique \cite{Sancho85jpf,MacKinnon85zpbc},
the conductance from the left terminal to the right terminal can be
evaluated as
\begin{equation}
G=\frac{e^{2}}{h}\mathrm{Tr}\left[\Gamma_{L}G^{r}\Gamma_{R}G^{a}\right],
\end{equation}
where $\Gamma_{L,R}$ are the line-width functions coupling to the
left lead and the right lead respectively, and $G^{r}(G^{a})$ is
the retarded (advanced) Green's function of the central region \cite{Datta}.

In the absence of a magnetic field, the value of two-terminal conductance
is exactly quantized at $G=2e^{2}/h$ as predicted theoretically for
the QSH effect. The conductance remains nearly unchanged for different
magnetic fields either along the boundary as shown in Fig.\ \ref{fig:conductance}(a)
or normal to the boundary as shown in Fig.\ \ref{fig:conductance}(b),
which can be attributed to the fact that the energy gap of edge states
is buried in the bulk valence bands. This support that the picture
of hidden Dirac point  may account for the experimental observations
on robust quantum edge transport in InAs/GaSb quantum wells \cite{DuLJ15prl}.
We also notice that a much stronger magnetic field makes the width
of the conductance plateau narrower, which indicates that the system
will be a semimetal under strong magnetic fields.

\section{Discussions and conclusions}

The gap opened in the edge states under an in-plane magnetic field
can be measured explicitly by means of reciprocal spin Hall effect
in a multi-terminal measurement \cite{Hankiewicz05prb}. The edge
state transport could survive even if the edge states and the bulk
electrons of valence bands co-exist, and can be checked in the non-local
measurement. This provides a possible way to verify the existence
of the edge states buried by the HH bands. However, the non-local
transport will disappear if the Fermi level sweeps over the energy
gap of the edge states in the presence of magnetic field if the bulk
electrons in the HH bands are presented.

In short, we re-examine the band structure and construct a six-band
effective model for InAs/GaSb quantum wells from the bulk Kane model.
An energy gap for helical edge states opens under a magnetic field,
which is well described by the effective g-factors of edge states.
The edge transport remains robust even though the magnetic field has
already broken time-reversal symmetry and opened an energy gap for
the helical edge states. This robustness is attributed to the peculiar
topological band structure that the Dirac point of the helical edge
states is buried in the bulk valence band after the anticrossing behavior.

\section{Acknowledgments}

C.L. and S.Z. thank Jia-Bin You, Jian Li and Lun-Hui Hu for helpful
discussions. This work was supported by the Research Grants Council,
University Grants Committee, Hong Kong under Grant No. 17304414 and
C6026-16W. HKU ITS computing facilities supported by the Hong Kong
UGC Special Equipment Grant (SEG HKU09).


\begin{thebibliography}{43}%
\makeatletter
\providecommand \@ifxundefined [1]{%
 \@ifx{#1\undefined}
}%
\providecommand \@ifnum [1]{%
 \ifnum #1\expandafter \@firstoftwo
 \else \expandafter \@secondoftwo
 \fi
}%
\providecommand \@ifx [1]{%
 \ifx #1\expandafter \@firstoftwo
 \else \expandafter \@secondoftwo
 \fi
}%
\providecommand \natexlab [1]{#1}%
\providecommand \enquote  [1]{``#1''}%
\providecommand \bibnamefont  [1]{#1}%
\providecommand \bibfnamefont [1]{#1}%
\providecommand \citenamefont [1]{#1}%
\providecommand \href@noop [0]{\@secondoftwo}%
\providecommand \href [0]{\begingroup \@sanitize@url \@href}%
\providecommand \@href[1]{\@@startlink{#1}\@@href}%
\providecommand \@@href[1]{\endgroup#1\@@endlink}%
\providecommand \@sanitize@url [0]{\catcode `\\12\catcode `\$12\catcode
  `\&12\catcode `\#12\catcode `\^12\catcode `\_12\catcode `\%12\relax}%
\providecommand \@@startlink[1]{}%
\providecommand \@@endlink[0]{}%
\providecommand \url  [0]{\begingroup\@sanitize@url \@url }%
\providecommand \@url [1]{\endgroup\@href {#1}{\urlprefix }}%
\providecommand \urlprefix  [0]{URL }%
\providecommand \Eprint [0]{\href }%
\providecommand \doibase [0]{http://dx.doi.org/}%
\providecommand \selectlanguage [0]{\@gobble}%
\providecommand \bibinfo  [0]{\@secondoftwo}%
\providecommand \bibfield  [0]{\@secondoftwo}%
\providecommand \translation [1]{[#1]}%
\providecommand \BibitemOpen [0]{}%
\providecommand \bibitemStop [0]{}%
\providecommand \bibitemNoStop [0]{.\EOS\space}%
\providecommand \EOS [0]{\spacefactor3000\relax}%
\providecommand \BibitemShut  [1]{\csname bibitem#1\endcsname}%
\let\auto@bib@innerbib\@empty
\bibitem [{\citenamefont {Hasan}\ and\ \citenamefont {Kane}(2010)}]{Kane10rmp}%
  \BibitemOpen
  \bibfield  {author} {\bibinfo {author} {\bibfnamefont {M.~Z.}\ \bibnamefont
  {Hasan}}\ and\ \bibinfo {author} {\bibfnamefont {C.~L.}\ \bibnamefont
  {Kane}},\ }\href {\doibase 10.1103/RevModPhys.82.3045} {\bibfield  {journal}
  {\bibinfo  {journal} {Rev. Mod. Phys.}\ }\textbf {\bibinfo {volume} {82}},\
  \bibinfo {pages} {3045} (\bibinfo {year} {2010})}\BibitemShut {NoStop}%
\bibitem [{\citenamefont {Qi}\ and\ \citenamefont {Zhang}(2011)}]{QiXL11rmp}%
  \BibitemOpen
  \bibfield  {author} {\bibinfo {author} {\bibfnamefont {X.-L.}\ \bibnamefont
  {Qi}}\ and\ \bibinfo {author} {\bibfnamefont {S.-C.}\ \bibnamefont {Zhang}},\
  }\href {\doibase 10.1103/RevModPhys.83.1057} {\bibfield  {journal} {\bibinfo
  {journal} {Rev. Mod. Phys.}\ }\textbf {\bibinfo {volume} {83}},\ \bibinfo
  {pages} {1057} (\bibinfo {year} {2011})}\BibitemShut {NoStop}%
\bibitem [{\citenamefont {Shen}(2017)}]{SQS}%
  \BibitemOpen
  \bibfield  {author} {\bibinfo {author} {\bibfnamefont {S.-Q.}\ \bibnamefont
  {Shen}},\ }\href@noop {} {\emph {\bibinfo {title} {Topological Insultaors:
  Dirac Equation in Condensed Matter}}},\ \bibinfo {edition} {2nd}\ ed.\
  (\bibinfo  {publisher} {Springer},\ \bibinfo {year} {2017})\BibitemShut
  {NoStop}%
\bibitem [{\citenamefont {Kane}\ and\ \citenamefont {Mele}(2005)}]{Kane05prlb}%
  \BibitemOpen
  \bibfield  {author} {\bibinfo {author} {\bibfnamefont {C.~L.}\ \bibnamefont
  {Kane}}\ and\ \bibinfo {author} {\bibfnamefont {E.~J.}\ \bibnamefont
  {Mele}},\ }\href {\doibase 10.1103/PhysRevLett.95.226801} {\bibfield
  {journal} {\bibinfo  {journal} {Phys. Rev. Lett.}\ }\textbf {\bibinfo
  {volume} {95}},\ \bibinfo {pages} {226801} (\bibinfo {year}
  {2005})}\BibitemShut {NoStop}%
\bibitem [{\citenamefont {Bernevig}\ \emph {et~al.}(2006)\citenamefont
  {Bernevig}, \citenamefont {Hughes},\ and\ \citenamefont
  {Zhang}}]{Bernevig06Scien}%
  \BibitemOpen
  \bibfield  {author} {\bibinfo {author} {\bibfnamefont {B.~A.}\ \bibnamefont
  {Bernevig}}, \bibinfo {author} {\bibfnamefont {T.~L.}\ \bibnamefont
  {Hughes}}, \ and\ \bibinfo {author} {\bibfnamefont {S.-C.}\ \bibnamefont
  {Zhang}},\ }\href {\doibase 10.1126/science.1133734} {\bibfield  {journal}
  {\bibinfo  {journal} {Science}\ }\textbf {\bibinfo {volume} {314}},\ \bibinfo
  {pages} {1757} (\bibinfo {year} {2006})}\BibitemShut {NoStop}%
\bibitem [{\citenamefont {K{\"o}nig}\ \emph {et~al.}(2007)\citenamefont
  {K{\"o}nig}, \citenamefont {Wiedmann}, \citenamefont {Br{\"u}ne},
  \citenamefont {Roth}, \citenamefont {Buhmann}, \citenamefont {Molenkamp},
  \citenamefont {Qi},\ and\ \citenamefont {Zhang}}]{Konig07Scien}%
  \BibitemOpen
  \bibfield  {author} {\bibinfo {author} {\bibfnamefont {M.}~\bibnamefont
  {K{\"o}nig}}, \bibinfo {author} {\bibfnamefont {S.}~\bibnamefont {Wiedmann}},
  \bibinfo {author} {\bibfnamefont {C.}~\bibnamefont {Br{\"u}ne}}, \bibinfo
  {author} {\bibfnamefont {A.}~\bibnamefont {Roth}}, \bibinfo {author}
  {\bibfnamefont {H.}~\bibnamefont {Buhmann}}, \bibinfo {author} {\bibfnamefont
  {L.~W.}\ \bibnamefont {Molenkamp}}, \bibinfo {author} {\bibfnamefont {X.-L.}\
  \bibnamefont {Qi}}, \ and\ \bibinfo {author} {\bibfnamefont {S.-C.}\
  \bibnamefont {Zhang}},\ }\href {\doibase 10.1126/science.1148047} {\bibfield
  {journal} {\bibinfo  {journal} {Science}\ }\textbf {\bibinfo {volume}
  {318}},\ \bibinfo {pages} {766} (\bibinfo {year} {2007})}\BibitemShut
  {NoStop}%
\bibitem [{\citenamefont {Roth}\ \emph {et~al.}(2009)\citenamefont {Roth},
  \citenamefont {Br{\"u}ne}, \citenamefont {Buhmann}, \citenamefont
  {Molenkamp}, \citenamefont {Maciejko}, \citenamefont {Qi},\ and\
  \citenamefont {Zhang}}]{Roth09Scien}%
  \BibitemOpen
  \bibfield  {author} {\bibinfo {author} {\bibfnamefont {A.}~\bibnamefont
  {Roth}}, \bibinfo {author} {\bibfnamefont {C.}~\bibnamefont {Br{\"u}ne}},
  \bibinfo {author} {\bibfnamefont {H.}~\bibnamefont {Buhmann}}, \bibinfo
  {author} {\bibfnamefont {L.~W.}\ \bibnamefont {Molenkamp}}, \bibinfo {author}
  {\bibfnamefont {J.}~\bibnamefont {Maciejko}}, \bibinfo {author}
  {\bibfnamefont {X.-L.}\ \bibnamefont {Qi}}, \ and\ \bibinfo {author}
  {\bibfnamefont {S.-C.}\ \bibnamefont {Zhang}},\ }\href {\doibase
  10.1126/science.1174736} {\bibfield  {journal} {\bibinfo  {journal}
  {Science}\ }\textbf {\bibinfo {volume} {325}},\ \bibinfo {pages} {294}
  (\bibinfo {year} {2009})}\BibitemShut {NoStop}%
\bibitem [{\citenamefont {Liu}\ \emph {et~al.}(2008)\citenamefont {Liu},
  \citenamefont {Hughes}, \citenamefont {Qi}, \citenamefont {Wang},\ and\
  \citenamefont {Zhang}}]{LiuCX08prl}%
  \BibitemOpen
  \bibfield  {author} {\bibinfo {author} {\bibfnamefont {C.}~\bibnamefont
  {Liu}}, \bibinfo {author} {\bibfnamefont {T.~L.}\ \bibnamefont {Hughes}},
  \bibinfo {author} {\bibfnamefont {X.-L.}\ \bibnamefont {Qi}}, \bibinfo
  {author} {\bibfnamefont {K.}~\bibnamefont {Wang}}, \ and\ \bibinfo {author}
  {\bibfnamefont {S.-C.}\ \bibnamefont {Zhang}},\ }\href {\doibase
  10.1103/PhysRevLett.100.236601} {\bibfield  {journal} {\bibinfo  {journal}
  {Phys. Rev. Lett.}\ }\textbf {\bibinfo {volume} {100}},\ \bibinfo {pages}
  {236601} (\bibinfo {year} {2008})}\BibitemShut {NoStop}%
\bibitem [{\citenamefont {Knez}\ \emph {et~al.}(2011)\citenamefont {Knez},
  \citenamefont {Du},\ and\ \citenamefont {Sullivan}}]{Knez11prl}%
  \BibitemOpen
  \bibfield  {author} {\bibinfo {author} {\bibfnamefont {I.}~\bibnamefont
  {Knez}}, \bibinfo {author} {\bibfnamefont {R.-R.}\ \bibnamefont {Du}}, \ and\
  \bibinfo {author} {\bibfnamefont {G.}~\bibnamefont {Sullivan}},\ }\href
  {\doibase 10.1103/PhysRevLett.107.136603} {\bibfield  {journal} {\bibinfo
  {journal} {Phys. Rev. Lett.}\ }\textbf {\bibinfo {volume} {107}},\ \bibinfo
  {pages} {136603} (\bibinfo {year} {2011})}\BibitemShut {NoStop}%
\bibitem [{\citenamefont {Knez}\ \emph {et~al.}(2010)\citenamefont {Knez},
  \citenamefont {Du},\ and\ \citenamefont {Sullivan}}]{Knez10prb}%
  \BibitemOpen
  \bibfield  {author} {\bibinfo {author} {\bibfnamefont {I.}~\bibnamefont
  {Knez}}, \bibinfo {author} {\bibfnamefont {R.~R.}\ \bibnamefont {Du}}, \ and\
  \bibinfo {author} {\bibfnamefont {G.}~\bibnamefont {Sullivan}},\ }\href
  {\doibase 10.1103/PhysRevB.81.201301} {\bibfield  {journal} {\bibinfo
  {journal} {Phys. Rev. B}\ }\textbf {\bibinfo {volume} {81}},\ \bibinfo
  {pages} {201301} (\bibinfo {year} {2010})}\BibitemShut {NoStop}%
\bibitem [{\citenamefont {Suzuki}\ \emph {et~al.}(2013)\citenamefont {Suzuki},
  \citenamefont {Harada}, \citenamefont {Onomitsu},\ and\ \citenamefont
  {Muraki}}]{Suzuki13prb}%
  \BibitemOpen
  \bibfield  {author} {\bibinfo {author} {\bibfnamefont {K.}~\bibnamefont
  {Suzuki}}, \bibinfo {author} {\bibfnamefont {Y.}~\bibnamefont {Harada}},
  \bibinfo {author} {\bibfnamefont {K.}~\bibnamefont {Onomitsu}}, \ and\
  \bibinfo {author} {\bibfnamefont {K.}~\bibnamefont {Muraki}},\ }\href
  {\doibase 10.1103/PhysRevB.87.235311} {\bibfield  {journal} {\bibinfo
  {journal} {Phys. Rev. B}\ }\textbf {\bibinfo {volume} {87}},\ \bibinfo
  {pages} {235311} (\bibinfo {year} {2013})}\BibitemShut {NoStop}%
\bibitem [{\citenamefont {Nichele}\ \emph {et~al.}(2014)\citenamefont
  {Nichele}, \citenamefont {Pal}, \citenamefont {Pietsch}, \citenamefont {Ihn},
  \citenamefont {Ensslin}, \citenamefont {Charpentier},\ and\ \citenamefont
  {Wegscheider}}]{Nichele14prl}%
  \BibitemOpen
  \bibfield  {author} {\bibinfo {author} {\bibfnamefont {F.}~\bibnamefont
  {Nichele}}, \bibinfo {author} {\bibfnamefont {A.~N.}\ \bibnamefont {Pal}},
  \bibinfo {author} {\bibfnamefont {P.}~\bibnamefont {Pietsch}}, \bibinfo
  {author} {\bibfnamefont {T.}~\bibnamefont {Ihn}}, \bibinfo {author}
  {\bibfnamefont {K.}~\bibnamefont {Ensslin}}, \bibinfo {author} {\bibfnamefont
  {C.}~\bibnamefont {Charpentier}}, \ and\ \bibinfo {author} {\bibfnamefont
  {W.}~\bibnamefont {Wegscheider}},\ }\href {\doibase
  10.1103/PhysRevLett.112.036802} {\bibfield  {journal} {\bibinfo  {journal}
  {Phys. Rev. Lett.}\ }\textbf {\bibinfo {volume} {112}},\ \bibinfo {pages}
  {036802} (\bibinfo {year} {2014})}\BibitemShut {NoStop}%
\bibitem [{\citenamefont {Mueller}\ \emph {et~al.}(2015)\citenamefont
  {Mueller}, \citenamefont {Pal}, \citenamefont {Karalic}, \citenamefont
  {Tschirky}, \citenamefont {Charpentier}, \citenamefont {Wegscheider},
  \citenamefont {Ensslin},\ and\ \citenamefont {Ihn}}]{Mueller15prb}%
  \BibitemOpen
  \bibfield  {author} {\bibinfo {author} {\bibfnamefont {S.}~\bibnamefont
  {Mueller}}, \bibinfo {author} {\bibfnamefont {A.~N.}\ \bibnamefont {Pal}},
  \bibinfo {author} {\bibfnamefont {M.}~\bibnamefont {Karalic}}, \bibinfo
  {author} {\bibfnamefont {T.}~\bibnamefont {Tschirky}}, \bibinfo {author}
  {\bibfnamefont {C.}~\bibnamefont {Charpentier}}, \bibinfo {author}
  {\bibfnamefont {W.}~\bibnamefont {Wegscheider}}, \bibinfo {author}
  {\bibfnamefont {K.}~\bibnamefont {Ensslin}}, \ and\ \bibinfo {author}
  {\bibfnamefont {T.}~\bibnamefont {Ihn}},\ }\href {\doibase
  10.1103/PhysRevB.92.081303} {\bibfield  {journal} {\bibinfo  {journal} {Phys.
  Rev. B}\ }\textbf {\bibinfo {volume} {92}},\ \bibinfo {pages} {081303}
  (\bibinfo {year} {2015})}\BibitemShut {NoStop}%
\bibitem [{\citenamefont {Qu}\ \emph {et~al.}(2015)\citenamefont {Qu},
  \citenamefont {Beukman}, \citenamefont {Nadj-Perge}, \citenamefont {Wimmer},
  \citenamefont {Nguyen}, \citenamefont {Yi}, \citenamefont {Thorp},
  \citenamefont {Sokolich}, \citenamefont {Kiselev}, \citenamefont {Manfra},
  \citenamefont {Marcus},\ and\ \citenamefont {Kouwenhoven}}]{QuF15prl}%
  \BibitemOpen
  \bibfield  {author} {\bibinfo {author} {\bibfnamefont {F.}~\bibnamefont
  {Qu}}, \bibinfo {author} {\bibfnamefont {A.~J.~A.}\ \bibnamefont {Beukman}},
  \bibinfo {author} {\bibfnamefont {S.}~\bibnamefont {Nadj-Perge}}, \bibinfo
  {author} {\bibfnamefont {M.}~\bibnamefont {Wimmer}}, \bibinfo {author}
  {\bibfnamefont {B.-M.}\ \bibnamefont {Nguyen}}, \bibinfo {author}
  {\bibfnamefont {W.}~\bibnamefont {Yi}}, \bibinfo {author} {\bibfnamefont
  {J.}~\bibnamefont {Thorp}}, \bibinfo {author} {\bibfnamefont
  {M.}~\bibnamefont {Sokolich}}, \bibinfo {author} {\bibfnamefont {A.~A.}\
  \bibnamefont {Kiselev}}, \bibinfo {author} {\bibfnamefont {M.~J.}\
  \bibnamefont {Manfra}}, \bibinfo {author} {\bibfnamefont {C.~M.}\
  \bibnamefont {Marcus}}, \ and\ \bibinfo {author} {\bibfnamefont {L.~P.}\
  \bibnamefont {Kouwenhoven}},\ }\href {\doibase
  10.1103/PhysRevLett.115.036803} {\bibfield  {journal} {\bibinfo  {journal}
  {Phys. Rev. Lett.}\ }\textbf {\bibinfo {volume} {115}},\ \bibinfo {pages}
  {036803} (\bibinfo {year} {2015})}\BibitemShut {NoStop}%
\bibitem [{\citenamefont {Nguyen}\ \emph {et~al.}(2016)\citenamefont {Nguyen},
  \citenamefont {Kiselev}, \citenamefont {Noah}, \citenamefont {Yi},
  \citenamefont {Qu}, \citenamefont {Beukman}, \citenamefont {de~Vries},
  \citenamefont {van Veen}, \citenamefont {Nadj-Perge}, \citenamefont
  {Kouwenhoven}, \citenamefont {Kjaergaard}, \citenamefont {Suominen},
  \citenamefont {Nichele}, \citenamefont {Marcus}, \citenamefont {Manfra},\
  and\ \citenamefont {Sokolich}}]{Nguyen16prl}%
  \BibitemOpen
  \bibfield  {author} {\bibinfo {author} {\bibfnamefont {B.-M.}\ \bibnamefont
  {Nguyen}}, \bibinfo {author} {\bibfnamefont {A.~A.}\ \bibnamefont {Kiselev}},
  \bibinfo {author} {\bibfnamefont {R.}~\bibnamefont {Noah}}, \bibinfo {author}
  {\bibfnamefont {W.}~\bibnamefont {Yi}}, \bibinfo {author} {\bibfnamefont
  {F.}~\bibnamefont {Qu}}, \bibinfo {author} {\bibfnamefont {A.~J.~A.}\
  \bibnamefont {Beukman}}, \bibinfo {author} {\bibfnamefont {F.~K.}\
  \bibnamefont {de~Vries}}, \bibinfo {author} {\bibfnamefont {J.}~\bibnamefont
  {van Veen}}, \bibinfo {author} {\bibfnamefont {S.}~\bibnamefont
  {Nadj-Perge}}, \bibinfo {author} {\bibfnamefont {L.~P.}\ \bibnamefont
  {Kouwenhoven}}, \bibinfo {author} {\bibfnamefont {M.}~\bibnamefont
  {Kjaergaard}}, \bibinfo {author} {\bibfnamefont {H.~J.}\ \bibnamefont
  {Suominen}}, \bibinfo {author} {\bibfnamefont {F.}~\bibnamefont {Nichele}},
  \bibinfo {author} {\bibfnamefont {C.~M.}\ \bibnamefont {Marcus}}, \bibinfo
  {author} {\bibfnamefont {M.~J.}\ \bibnamefont {Manfra}}, \ and\ \bibinfo
  {author} {\bibfnamefont {M.}~\bibnamefont {Sokolich}},\ }\href {\doibase
  10.1103/PhysRevLett.117.077701} {\bibfield  {journal} {\bibinfo  {journal}
  {Phys. Rev. Lett.}\ }\textbf {\bibinfo {volume} {117}},\ \bibinfo {pages}
  {077701} (\bibinfo {year} {2016})}\BibitemShut {NoStop}%
\bibitem [{\citenamefont {Karalic}\ \emph {et~al.}(2016)\citenamefont
  {Karalic}, \citenamefont {Mueller}, \citenamefont {Mittag}, \citenamefont
  {Pakrouski}, \citenamefont {Wu}, \citenamefont {Soluyanov}, \citenamefont
  {Troyer}, \citenamefont {Tschirky}, \citenamefont {Wegscheider},
  \citenamefont {Ensslin},\ and\ \citenamefont {Ihn}}]{Karalic16prbR}%
  \BibitemOpen
  \bibfield  {author} {\bibinfo {author} {\bibfnamefont {M.}~\bibnamefont
  {Karalic}}, \bibinfo {author} {\bibfnamefont {S.}~\bibnamefont {Mueller}},
  \bibinfo {author} {\bibfnamefont {C.}~\bibnamefont {Mittag}}, \bibinfo
  {author} {\bibfnamefont {K.}~\bibnamefont {Pakrouski}}, \bibinfo {author}
  {\bibfnamefont {Q.}~\bibnamefont {Wu}}, \bibinfo {author} {\bibfnamefont
  {A.~A.}\ \bibnamefont {Soluyanov}}, \bibinfo {author} {\bibfnamefont
  {M.}~\bibnamefont {Troyer}}, \bibinfo {author} {\bibfnamefont
  {T.}~\bibnamefont {Tschirky}}, \bibinfo {author} {\bibfnamefont
  {W.}~\bibnamefont {Wegscheider}}, \bibinfo {author} {\bibfnamefont
  {K.}~\bibnamefont {Ensslin}}, \ and\ \bibinfo {author} {\bibfnamefont
  {T.}~\bibnamefont {Ihn}},\ }\href {\doibase 10.1103/PhysRevB.94.241402}
  {\bibfield  {journal} {\bibinfo  {journal} {Phys. Rev. B}\ }\textbf {\bibinfo
  {volume} {94}},\ \bibinfo {pages} {241402} (\bibinfo {year}
  {2016})}\BibitemShut {NoStop}%
\bibitem [{\citenamefont {Nichele}\ \emph {et~al.}(2017)\citenamefont
  {Nichele}, \citenamefont {Kjaergaard}, \citenamefont {Suominen},
  \citenamefont {Skolasinski}, \citenamefont {Wimmer}, \citenamefont {Nguyen},
  \citenamefont {Kiselev}, \citenamefont {Yi}, \citenamefont {Sokolich},
  \citenamefont {Manfra}, \citenamefont {Qu}, \citenamefont {Beukman},
  \citenamefont {Kouwenhoven},\ and\ \citenamefont {Marcus}}]{Nichele17prl}%
  \BibitemOpen
  \bibfield  {author} {\bibinfo {author} {\bibfnamefont {F.}~\bibnamefont
  {Nichele}}, \bibinfo {author} {\bibfnamefont {M.}~\bibnamefont {Kjaergaard}},
  \bibinfo {author} {\bibfnamefont {H.~J.}\ \bibnamefont {Suominen}}, \bibinfo
  {author} {\bibfnamefont {R.}~\bibnamefont {Skolasinski}}, \bibinfo {author}
  {\bibfnamefont {M.}~\bibnamefont {Wimmer}}, \bibinfo {author} {\bibfnamefont
  {B.-M.}\ \bibnamefont {Nguyen}}, \bibinfo {author} {\bibfnamefont {A.~A.}\
  \bibnamefont {Kiselev}}, \bibinfo {author} {\bibfnamefont {W.}~\bibnamefont
  {Yi}}, \bibinfo {author} {\bibfnamefont {M.}~\bibnamefont {Sokolich}},
  \bibinfo {author} {\bibfnamefont {M.~J.}\ \bibnamefont {Manfra}}, \bibinfo
  {author} {\bibfnamefont {F.}~\bibnamefont {Qu}}, \bibinfo {author}
  {\bibfnamefont {A.~J.~A.}\ \bibnamefont {Beukman}}, \bibinfo {author}
  {\bibfnamefont {L.~P.}\ \bibnamefont {Kouwenhoven}}, \ and\ \bibinfo {author}
  {\bibfnamefont {C.~M.}\ \bibnamefont {Marcus}},\ }\href {\doibase
  10.1103/PhysRevLett.118.016801} {\bibfield  {journal} {\bibinfo  {journal}
  {Phys. Rev. Lett.}\ }\textbf {\bibinfo {volume} {118}},\ \bibinfo {pages}
  {016801} (\bibinfo {year} {2017})}\BibitemShut {NoStop}%
\bibitem [{\citenamefont {Du}\ \emph {et~al.}(2015)\citenamefont {Du},
  \citenamefont {Knez}, \citenamefont {Sullivan},\ and\ \citenamefont
  {Du}}]{DuLJ15prl}%
  \BibitemOpen
  \bibfield  {author} {\bibinfo {author} {\bibfnamefont {L.}~\bibnamefont
  {Du}}, \bibinfo {author} {\bibfnamefont {I.}~\bibnamefont {Knez}}, \bibinfo
  {author} {\bibfnamefont {G.}~\bibnamefont {Sullivan}}, \ and\ \bibinfo
  {author} {\bibfnamefont {R.-R.}\ \bibnamefont {Du}},\ }\href {\doibase
  10.1103/PhysRevLett.114.096802} {\bibfield  {journal} {\bibinfo  {journal}
  {Phys. Rev. Lett.}\ }\textbf {\bibinfo {volume} {114}},\ \bibinfo {pages}
  {096802} (\bibinfo {year} {2015})}\BibitemShut {NoStop}%
\bibitem [{\citenamefont {Ma}\ \emph {et~al.}(2015)\citenamefont {Ma},
  \citenamefont {Calvo}, \citenamefont {Wang}, \citenamefont {Lian},
  \citenamefont {M\"uhlbauer}, \citenamefont {Br\"une}, \citenamefont {Cui},
  \citenamefont {Lai}, \citenamefont {Kundhikanjana}, \citenamefont {Yang},
  \citenamefont {Baenninger}, \citenamefont {K\"onig}, \citenamefont {Ames},
  \citenamefont {Buhmann}, \citenamefont {Leubner}, \citenamefont {Molenkamp},
  \citenamefont {Zhang}, \citenamefont {Goldhaber-Gordon}, \citenamefont
  {Kelly},\ and\ \citenamefont {Shen}}]{Ma15natcomm}%
  \BibitemOpen
  \bibfield  {author} {\bibinfo {author} {\bibfnamefont {E.~Y.}\ \bibnamefont
  {Ma}}, \bibinfo {author} {\bibfnamefont {M.~R.}\ \bibnamefont {Calvo}},
  \bibinfo {author} {\bibfnamefont {J.}~\bibnamefont {Wang}}, \bibinfo {author}
  {\bibfnamefont {B.}~\bibnamefont {Lian}}, \bibinfo {author} {\bibfnamefont
  {M.}~\bibnamefont {M\"uhlbauer}}, \bibinfo {author} {\bibfnamefont
  {C.}~\bibnamefont {Br\"une}}, \bibinfo {author} {\bibfnamefont {Y.-T.}\
  \bibnamefont {Cui}}, \bibinfo {author} {\bibfnamefont {K.}~\bibnamefont
  {Lai}}, \bibinfo {author} {\bibfnamefont {W.}~\bibnamefont {Kundhikanjana}},
  \bibinfo {author} {\bibfnamefont {Y.}~\bibnamefont {Yang}}, \bibinfo {author}
  {\bibfnamefont {M.}~\bibnamefont {Baenninger}}, \bibinfo {author}
  {\bibfnamefont {M.}~\bibnamefont {K\"onig}}, \bibinfo {author} {\bibfnamefont
  {C.}~\bibnamefont {Ames}}, \bibinfo {author} {\bibfnamefont {H.}~\bibnamefont
  {Buhmann}}, \bibinfo {author} {\bibfnamefont {P.}~\bibnamefont {Leubner}},
  \bibinfo {author} {\bibfnamefont {L.~W.}\ \bibnamefont {Molenkamp}}, \bibinfo
  {author} {\bibfnamefont {S.-C.}\ \bibnamefont {Zhang}}, \bibinfo {author}
  {\bibfnamefont {D.}~\bibnamefont {Goldhaber-Gordon}}, \bibinfo {author}
  {\bibfnamefont {M.~A.}\ \bibnamefont {Kelly}}, \ and\ \bibinfo {author}
  {\bibfnamefont {Z.-X.}\ \bibnamefont {Shen}},\ }\href
  {http://dx.doi.org/10.1038/ncomms8252} {\bibfield  {journal} {\bibinfo
  {journal} {Nat. Commun.}\ }\textbf {\bibinfo {volume} {6}},\ \bibinfo {pages}
  {7252} (\bibinfo {year} {2015})}\BibitemShut {NoStop}%
\bibitem [{\citenamefont {Pikulin}\ \emph {et~al.}(2014)\citenamefont
  {Pikulin}, \citenamefont {Hyart}, \citenamefont {Mi}, \citenamefont
  {Tworzyd\l{}o}, \citenamefont {Wimmer},\ and\ \citenamefont
  {Beenakker}}]{Pikulin14prb}%
  \BibitemOpen
  \bibfield  {author} {\bibinfo {author} {\bibfnamefont {D.~I.}\ \bibnamefont
  {Pikulin}}, \bibinfo {author} {\bibfnamefont {T.}~\bibnamefont {Hyart}},
  \bibinfo {author} {\bibfnamefont {S.}~\bibnamefont {Mi}}, \bibinfo {author}
  {\bibfnamefont {J.}~\bibnamefont {Tworzyd\l{}o}}, \bibinfo {author}
  {\bibfnamefont {M.}~\bibnamefont {Wimmer}}, \ and\ \bibinfo {author}
  {\bibfnamefont {C.~W.~J.}\ \bibnamefont {Beenakker}},\ }\href {\doibase
  10.1103/PhysRevB.89.161403} {\bibfield  {journal} {\bibinfo  {journal} {Phys.
  Rev. B}\ }\textbf {\bibinfo {volume} {89}},\ \bibinfo {pages} {161403}
  (\bibinfo {year} {2014})}\BibitemShut {NoStop}%
\bibitem [{\citenamefont {Zhang}\ \emph {et~al.}(2014)\citenamefont {Zhang},
  \citenamefont {Zhang},\ and\ \citenamefont {Shen}}]{ZhangSB14prb}%
  \BibitemOpen
  \bibfield  {author} {\bibinfo {author} {\bibfnamefont {S.-B.}\ \bibnamefont
  {Zhang}}, \bibinfo {author} {\bibfnamefont {Y.-Y.}\ \bibnamefont {Zhang}}, \
  and\ \bibinfo {author} {\bibfnamefont {S.-Q.}\ \bibnamefont {Shen}},\ }\href
  {\doibase 10.1103/PhysRevB.90.115305} {\bibfield  {journal} {\bibinfo
  {journal} {Phys. Rev. B}\ }\textbf {\bibinfo {volume} {90}},\ \bibinfo
  {pages} {115305} (\bibinfo {year} {2014})}\BibitemShut {NoStop}%
\bibitem [{\citenamefont {Hu}\ \emph {et~al.}(2016)\citenamefont {Hu},
  \citenamefont {Xu}, \citenamefont {Zhang},\ and\ \citenamefont
  {Zhou}}]{HuLH16prb}%
  \BibitemOpen
  \bibfield  {author} {\bibinfo {author} {\bibfnamefont {L.-H.}\ \bibnamefont
  {Hu}}, \bibinfo {author} {\bibfnamefont {D.-H.}\ \bibnamefont {Xu}}, \bibinfo
  {author} {\bibfnamefont {F.-C.}\ \bibnamefont {Zhang}}, \ and\ \bibinfo
  {author} {\bibfnamefont {Y.}~\bibnamefont {Zhou}},\ }\href {\doibase
  10.1103/PhysRevB.94.085306} {\bibfield  {journal} {\bibinfo  {journal} {Phys.
  Rev. B}\ }\textbf {\bibinfo {volume} {94}},\ \bibinfo {pages} {085306}
  (\bibinfo {year} {2016})}\BibitemShut {NoStop}%
\bibitem [{\citenamefont {Altarelli}(1983)}]{Altarelli83prb}%
  \BibitemOpen
  \bibfield  {author} {\bibinfo {author} {\bibfnamefont {M.}~\bibnamefont
  {Altarelli}},\ }\href {\doibase 10.1103/PhysRevB.28.842} {\bibfield
  {journal} {\bibinfo  {journal} {Phys. Rev. B}\ }\textbf {\bibinfo {volume}
  {28}},\ \bibinfo {pages} {842} (\bibinfo {year} {1983})}\BibitemShut
  {NoStop}%
\bibitem [{\citenamefont {Yang}\ \emph {et~al.}(1997)\citenamefont {Yang},
  \citenamefont {Yang}, \citenamefont {Bennett},\ and\ \citenamefont
  {Shanabrook}}]{YangM97prl}%
  \BibitemOpen
  \bibfield  {author} {\bibinfo {author} {\bibfnamefont {M.~J.}\ \bibnamefont
  {Yang}}, \bibinfo {author} {\bibfnamefont {C.~H.}\ \bibnamefont {Yang}},
  \bibinfo {author} {\bibfnamefont {B.~R.}\ \bibnamefont {Bennett}}, \ and\
  \bibinfo {author} {\bibfnamefont {B.~V.}\ \bibnamefont {Shanabrook}},\ }\href
  {\doibase 10.1103/PhysRevLett.78.4613} {\bibfield  {journal} {\bibinfo
  {journal} {Phys. Rev. Lett.}\ }\textbf {\bibinfo {volume} {78}},\ \bibinfo
  {pages} {4613} (\bibinfo {year} {1997})}\BibitemShut {NoStop}%
\bibitem [{\citenamefont {Naveh}\ and\ \citenamefont
  {Laikhtman}(1995)}]{Naveh95apl}%
  \BibitemOpen
  \bibfield  {author} {\bibinfo {author} {\bibfnamefont {Y.}~\bibnamefont
  {Naveh}}\ and\ \bibinfo {author} {\bibfnamefont {B.}~\bibnamefont
  {Laikhtman}},\ }\href {\doibase 10.1063/1.113297} {\bibfield  {journal}
  {\bibinfo  {journal} {App. Phys. Lett.}\ }\textbf {\bibinfo {volume} {66}},\
  \bibinfo {pages} {1980} (\bibinfo {year} {1995})}\BibitemShut {NoStop}%
\bibitem [{\citenamefont {Kane}(1957)}]{Kane57JPCS}%
  \BibitemOpen
  \bibfield  {author} {\bibinfo {author} {\bibfnamefont {E.~O.}\ \bibnamefont
  {Kane}},\ }\href {\doibase http://dx.doi.org/10.1016/0022-3697(57)90013-6}
  {\bibfield  {journal} {\bibinfo  {journal} {J. Phys. Chem. Solids}\ }\textbf
  {\bibinfo {volume} {1}},\ \bibinfo {pages} {249 } (\bibinfo {year}
  {1957})}\BibitemShut {NoStop}%
\bibitem [{\citenamefont {Winkler}(2003)}]{Winkler}%
  \BibitemOpen
  \bibfield  {author} {\bibinfo {author} {\bibfnamefont {R.}~\bibnamefont
  {Winkler}},\ }\href@noop {} {\emph {\bibinfo {title} {Spin-orbit Coupling in
  Two-Dimensional Electron and Hole Systems}}}\ (\bibinfo  {publisher}
  {Springer-Verlag, Berlin},\ \bibinfo {year} {2003})\BibitemShut {NoStop}%
\bibitem [{\citenamefont {Novik}\ \emph {et~al.}(2005)\citenamefont {Novik},
  \citenamefont {Pfeuffer-Jeschke}, \citenamefont {Jungwirth}, \citenamefont
  {Latussek}, \citenamefont {Becker}, \citenamefont {Landwehr}, \citenamefont
  {Buhmann},\ and\ \citenamefont {Molenkamp}}]{Novik05prb}%
  \BibitemOpen
  \bibfield  {author} {\bibinfo {author} {\bibfnamefont {E.~G.}\ \bibnamefont
  {Novik}}, \bibinfo {author} {\bibfnamefont {A.}~\bibnamefont
  {Pfeuffer-Jeschke}}, \bibinfo {author} {\bibfnamefont {T.}~\bibnamefont
  {Jungwirth}}, \bibinfo {author} {\bibfnamefont {V.}~\bibnamefont {Latussek}},
  \bibinfo {author} {\bibfnamefont {C.~R.}\ \bibnamefont {Becker}}, \bibinfo
  {author} {\bibfnamefont {G.}~\bibnamefont {Landwehr}}, \bibinfo {author}
  {\bibfnamefont {H.}~\bibnamefont {Buhmann}}, \ and\ \bibinfo {author}
  {\bibfnamefont {L.~W.}\ \bibnamefont {Molenkamp}},\ }\href {\doibase
  10.1103/PhysRevB.72.035321} {\bibfield  {journal} {\bibinfo  {journal} {Phys.
  Rev. B}\ }\textbf {\bibinfo {volume} {72}},\ \bibinfo {pages} {035321}
  (\bibinfo {year} {2005})}\BibitemShut {NoStop}%
\bibitem [{\citenamefont {Lawaetz}(1971)}]{Lawaetz71prb}%
  \BibitemOpen
  \bibfield  {author} {\bibinfo {author} {\bibfnamefont {P.}~\bibnamefont
  {Lawaetz}},\ }\href {\doibase 10.1103/PhysRevB.4.3460} {\bibfield  {journal}
  {\bibinfo  {journal} {Phys. Rev. B}\ }\textbf {\bibinfo {volume} {4}},\
  \bibinfo {pages} {3460} (\bibinfo {year} {1971})}\BibitemShut {NoStop}%
\bibitem [{\citenamefont {Halvorsen}\ \emph {et~al.}(2000)\citenamefont
  {Halvorsen}, \citenamefont {Galperin},\ and\ \citenamefont
  {Chao}}]{Halvorsen99prb}%
  \BibitemOpen
  \bibfield  {author} {\bibinfo {author} {\bibfnamefont {E.}~\bibnamefont
  {Halvorsen}}, \bibinfo {author} {\bibfnamefont {Y.}~\bibnamefont {Galperin}},
  \ and\ \bibinfo {author} {\bibfnamefont {K.~A.}\ \bibnamefont {Chao}},\
  }\href {\doibase 10.1103/PhysRevB.61.16743} {\bibfield  {journal} {\bibinfo
  {journal} {Phys. Rev. B}\ }\textbf {\bibinfo {volume} {61}},\ \bibinfo
  {pages} {16743} (\bibinfo {year} {2000})}\BibitemShut {NoStop}%
\bibitem [{\citenamefont {Li}\ \emph {et~al.}(2009)\citenamefont {Li},
  \citenamefont {Yang},\ and\ \citenamefont {Chang}}]{LiJ09prb}%
  \BibitemOpen
  \bibfield  {author} {\bibinfo {author} {\bibfnamefont {J.}~\bibnamefont
  {Li}}, \bibinfo {author} {\bibfnamefont {W.}~\bibnamefont {Yang}}, \ and\
  \bibinfo {author} {\bibfnamefont {K.}~\bibnamefont {Chang}},\ }\href
  {\doibase 10.1103/PhysRevB.80.035303} {\bibfield  {journal} {\bibinfo
  {journal} {Phys. Rev. B}\ }\textbf {\bibinfo {volume} {80}},\ \bibinfo
  {pages} {035303} (\bibinfo {year} {2009})}\BibitemShut {NoStop}%
\bibitem [{\citenamefont {Hatsugai}(1993)}]{Hatsugai93prl}%
  \BibitemOpen
  \bibfield  {author} {\bibinfo {author} {\bibfnamefont {Y.}~\bibnamefont
  {Hatsugai}},\ }\href {\doibase 10.1103/PhysRevLett.71.3697} {\bibfield
  {journal} {\bibinfo  {journal} {Phys. Rev. Lett.}\ }\textbf {\bibinfo
  {volume} {71}},\ \bibinfo {pages} {3697} (\bibinfo {year}
  {1993})}\BibitemShut {NoStop}%
\bibitem [{\citenamefont {Qi}\ \emph {et~al.}(2006)\citenamefont {Qi},
  \citenamefont {Wu},\ and\ \citenamefont {Zhang}}]{Qi06prb}%
  \BibitemOpen
  \bibfield  {author} {\bibinfo {author} {\bibfnamefont {X.-L.}\ \bibnamefont
  {Qi}}, \bibinfo {author} {\bibfnamefont {Y.-S.}\ \bibnamefont {Wu}}, \ and\
  \bibinfo {author} {\bibfnamefont {S.-C.}\ \bibnamefont {Zhang}},\ }\href
  {\doibase 10.1103/PhysRevB.74.045125} {\bibfield  {journal} {\bibinfo
  {journal} {Phys. Rev. B}\ }\textbf {\bibinfo {volume} {74}},\ \bibinfo
  {pages} {045125} (\bibinfo {year} {2006})}\BibitemShut {NoStop}%
\bibitem [{\citenamefont {Graf}\ and\ \citenamefont {Porta}(2013)}]{Graf2013}%
  \BibitemOpen
  \bibfield  {author} {\bibinfo {author} {\bibfnamefont {G.~M.}\ \bibnamefont
  {Graf}}\ and\ \bibinfo {author} {\bibfnamefont {M.}~\bibnamefont {Porta}},\
  }\href {\doibase 10.1007/s00220-013-1819-6} {\bibfield  {journal} {\bibinfo
  {journal} {Commun. Math. Phys.}\ }\textbf {\bibinfo {volume} {324}},\
  \bibinfo {pages} {851} (\bibinfo {year} {2013})}\BibitemShut {NoStop}%
\bibitem [{\citenamefont {Rothe}\ \emph {et~al.}(2010)\citenamefont {Rothe},
  \citenamefont {Reinthaler}, \citenamefont {Liu}, \citenamefont {Molenkamp},
  \citenamefont {Zhang},\ and\ \citenamefont {Hankiewicz}}]{Roth10njp}%
  \BibitemOpen
  \bibfield  {author} {\bibinfo {author} {\bibfnamefont {D.~G.}\ \bibnamefont
  {Rothe}}, \bibinfo {author} {\bibfnamefont {R.~W.}\ \bibnamefont
  {Reinthaler}}, \bibinfo {author} {\bibfnamefont {C.-X.}\ \bibnamefont {Liu}},
  \bibinfo {author} {\bibfnamefont {L.~W.}\ \bibnamefont {Molenkamp}}, \bibinfo
  {author} {\bibfnamefont {S.-C.}\ \bibnamefont {Zhang}}, \ and\ \bibinfo
  {author} {\bibfnamefont {E.~M.}\ \bibnamefont {Hankiewicz}},\ }\href
  {http://stacks.iop.org/1367-2630/12/i=6/a=065012} {\bibfield  {journal}
  {\bibinfo  {journal} {New J. Phys.}\ }\textbf {\bibinfo {volume} {12}},\
  \bibinfo {pages} {065012} (\bibinfo {year} {2010})}\BibitemShut {NoStop}%
\bibitem [{\citenamefont {Beugeling}\ \emph {et~al.}(2012)\citenamefont
  {Beugeling}, \citenamefont {Liu}, \citenamefont {Novik}, \citenamefont
  {Molenkamp},\ and\ \citenamefont {Morais~Smith}}]{Beugeling12prb}%
  \BibitemOpen
  \bibfield  {author} {\bibinfo {author} {\bibfnamefont {W.}~\bibnamefont
  {Beugeling}}, \bibinfo {author} {\bibfnamefont {C.~X.}\ \bibnamefont {Liu}},
  \bibinfo {author} {\bibfnamefont {E.~G.}\ \bibnamefont {Novik}}, \bibinfo
  {author} {\bibfnamefont {L.~W.}\ \bibnamefont {Molenkamp}}, \ and\ \bibinfo
  {author} {\bibfnamefont {C.}~\bibnamefont {Morais~Smith}},\ }\href {\doibase
  10.1103/PhysRevB.85.195304} {\bibfield  {journal} {\bibinfo  {journal} {Phys.
  Rev. B}\ }\textbf {\bibinfo {volume} {85}},\ \bibinfo {pages} {195304}
  (\bibinfo {year} {2012})}\BibitemShut {NoStop}%
\bibitem [{\citenamefont {Zakharova}\ \emph {et~al.}(2003)\citenamefont
  {Zakharova}, \citenamefont {Yen},\ and\ \citenamefont {Chao}}]{ChaoKA03jn}%
  \BibitemOpen
  \bibfield  {author} {\bibinfo {author} {\bibfnamefont {A.}~\bibnamefont
  {Zakharova}}, \bibinfo {author} {\bibfnamefont {S.~T.}\ \bibnamefont {Yen}},
  \ and\ \bibinfo {author} {\bibfnamefont {K.~A.}\ \bibnamefont {Chao}},\
  }\href {\doibase 10.1142/S0219581X0300153X} {\bibfield  {journal} {\bibinfo
  {journal} {Int. J. Nanosci.}\ }\textbf {\bibinfo {volume} {02}},\ \bibinfo
  {pages} {437} (\bibinfo {year} {2003})}\BibitemShut {NoStop}%
\bibitem [{\citenamefont {Mu}\ \emph {et~al.}(2016)\citenamefont {Mu},
  \citenamefont {Sullivan},\ and\ \citenamefont {Du}}]{MuXY16apl}%
  \BibitemOpen
  \bibfield  {author} {\bibinfo {author} {\bibfnamefont {X.}~\bibnamefont
  {Mu}}, \bibinfo {author} {\bibfnamefont {G.}~\bibnamefont {Sullivan}}, \ and\
  \bibinfo {author} {\bibfnamefont {R.-R.}\ \bibnamefont {Du}},\ }\href
  {http://scitation.aip.org/content/aip/journal/apl/108/1/10.1063/1.4939230}
  {\bibfield  {journal} {\bibinfo  {journal} {App. Phys. Lett.}\ }\textbf
  {\bibinfo {volume} {108}},\ \bibinfo {eid} {012101} (\bibinfo {year}
  {2016})}\BibitemShut {NoStop}%
\bibitem [{\citenamefont {Yang}\ \emph {et~al.}(2011)\citenamefont {Yang},
  \citenamefont {Xu}, \citenamefont {Sheng}, \citenamefont {Wang},
  \citenamefont {Xing},\ and\ \citenamefont {Sheng}}]{YangY11prl}%
  \BibitemOpen
  \bibfield  {author} {\bibinfo {author} {\bibfnamefont {Y.}~\bibnamefont
  {Yang}}, \bibinfo {author} {\bibfnamefont {Z.}~\bibnamefont {Xu}}, \bibinfo
  {author} {\bibfnamefont {L.}~\bibnamefont {Sheng}}, \bibinfo {author}
  {\bibfnamefont {B.}~\bibnamefont {Wang}}, \bibinfo {author} {\bibfnamefont
  {D.~Y.}\ \bibnamefont {Xing}}, \ and\ \bibinfo {author} {\bibfnamefont
  {D.~N.}\ \bibnamefont {Sheng}},\ }\href {\doibase
  10.1103/PhysRevLett.107.066602} {\bibfield  {journal} {\bibinfo  {journal}
  {Phys. Rev. Lett.}\ }\textbf {\bibinfo {volume} {107}},\ \bibinfo {pages}
  {066602} (\bibinfo {year} {2011})}\BibitemShut {NoStop}%
\bibitem [{\citenamefont {Sancho}\ \emph {et~al.}(1985)\citenamefont {Sancho},
  \citenamefont {Sancho}, \citenamefont {Sancho},\ and\ \citenamefont
  {Rubio}}]{Sancho85jpf}%
  \BibitemOpen
  \bibfield  {author} {\bibinfo {author} {\bibfnamefont {M.~P.~L.}\
  \bibnamefont {Sancho}}, \bibinfo {author} {\bibfnamefont {J.~M.~L.}\
  \bibnamefont {Sancho}}, \bibinfo {author} {\bibfnamefont {J.~M.~L.}\
  \bibnamefont {Sancho}}, \ and\ \bibinfo {author} {\bibfnamefont
  {J.}~\bibnamefont {Rubio}},\ }\href@noop {} {\bibfield  {journal} {\bibinfo
  {journal} {Journal of Physics F: Metal Physics}\ }\textbf {\bibinfo {volume}
  {15}},\ \bibinfo {pages} {851} (\bibinfo {year} {1985})}\BibitemShut
  {NoStop}%
\bibitem [{\citenamefont {MacKinnon}(1985)}]{MacKinnon85zpbc}%
  \BibitemOpen
  \bibfield  {author} {\bibinfo {author} {\bibfnamefont {A.}~\bibnamefont
  {MacKinnon}},\ }\href {\doibase 10.1007/BF01328846} {\bibfield  {journal}
  {\bibinfo  {journal} {Z. Phys. B Condensed Matter}\ }\textbf {\bibinfo
  {volume} {59}},\ \bibinfo {pages} {385} (\bibinfo {year} {1985})}\BibitemShut
  {NoStop}%
\bibitem [{\citenamefont {Datta}(1995)}]{Datta}%
  \BibitemOpen
  \bibfield  {author} {\bibinfo {author} {\bibfnamefont {S.}~\bibnamefont
  {Datta}},\ }\href@noop {} {\emph {\bibinfo {title} {Electronic Transport in
  Mesoscopic Systems}}}\ (\bibinfo  {publisher} {Cambridge University Press},\
  \bibinfo {year} {1995})\BibitemShut {NoStop}%
\bibitem [{\citenamefont {Hankiewicz}\ \emph {et~al.}(2005)\citenamefont
  {Hankiewicz}, \citenamefont {Li}, \citenamefont {Jungwirth}, \citenamefont
  {Niu}, \citenamefont {Shen},\ and\ \citenamefont {Sinova}}]{Hankiewicz05prb}%
  \BibitemOpen
  \bibfield  {author} {\bibinfo {author} {\bibfnamefont {E.~M.}\ \bibnamefont
  {Hankiewicz}}, \bibinfo {author} {\bibfnamefont {J.}~\bibnamefont {Li}},
  \bibinfo {author} {\bibfnamefont {T.}~\bibnamefont {Jungwirth}}, \bibinfo
  {author} {\bibfnamefont {Q.}~\bibnamefont {Niu}}, \bibinfo {author}
  {\bibfnamefont {S.-Q.}\ \bibnamefont {Shen}}, \ and\ \bibinfo {author}
  {\bibfnamefont {J.}~\bibnamefont {Sinova}},\ }\href {\doibase
  10.1103/PhysRevB.72.155305} {\bibfield  {journal} {\bibinfo  {journal} {Phys.
  Rev. B}\ }\textbf {\bibinfo {volume} {72}},\ \bibinfo {pages} {155305}
  (\bibinfo {year} {2005})}\BibitemShut {NoStop}%
\end{thebibliography}

%

\end{document}